\documentclass[fleqn,usenatbib]{mnras}

\usepackage{newtxtext,newtxmath}

\usepackage[T1]{fontenc}
\usepackage{tablefootnote}
\usepackage{natbib}
\defcitealias{Collins2013}{C13}
\defcitealias{Read2019}{R19}

\DeclareRobustCommand{\VAN}[3]{#2}
\let\VANthebibliography\thebibliography
\def\thebibliography{\DeclareRobustCommand{\VAN}[3]{##3}\VANthebibliography}

\usepackage{graphicx}	
\usepackage{amsmath}	

\title[Andromeda XXV in a low density halo]{Andromeda XXV - a dwarf galaxy with a low central dark matter density}

\author[E. J. E. Charles et al.]{Emily J. E. Charles,$^{1}$\thanks{E-mail: e.charles@surrey.ac.uk}
Michelle L. M. Collins,$^{1}$
R. Michael Rich,$^{2}$
Justin I. Read,$^{1}$
Stacy Y. Kim,$^{1}$
\newauthor Rodrigo A. Ibata$^{3}$
Nicolas F. Martin,$^{3}$
Scott C. Chapman,$^{4}$
Eduardo Balbinot,$^{5}$
Daniel R. Weisz $^{6}$
\\
$^{1}$Physics Department, University of Surrey, Guildford, GU2 7XH, UK\\
$^{2}$Department of Physics and Astronomy, University of California at Los Angeles, Los Angeles, CA 90095, USA\\
$^{3}$Observatoire astronomique de Strasbourg, Universi\'e de Strasbourg, CNRS, UMR 7550, 11 rue de l’Universit\'e, F-67000 Strasbourg, France\\
$^{4}$Department of Physics and Atmospheric Science, Dalhousie University, Coburg Road, Halifax B3H1A6, Canada\\
$^{5}$Kapteyn Astronomical Institute, University of Groningen, Landleven 12, NL-9747 AD Groningen, the Netherlands\\
$^{6}$ Department of Astronomy, University of California Berkeley, Berkeley, CA 94720, USA\\
}

\date{Accepted XXX. Received YYY; in original form ZZZ}

\pubyear{2022}

\begin{document}
\label{firstpage}
\pagerange{\pageref{firstpage}--\pageref{lastpage}}
\maketitle

\begin{abstract}
Andromeda (And) XXV has previously been reported as a dwarf spheroidal galaxy (dSph) with little-to-no dark matter. However, the uncertainties on this result were significant. In this study, we double the number of member stars and re-derive the kinematics and mass of And XXV. We find that And~XXV has a systemic velocity of $\nu_{\rm r}=-107.7\pm1.0$~kms$^{-1}$ and a velocity dispersion of $\sigma_{\rm \nu}=4.5\pm1.0$~kms$^{-1}$. With this better constrained velocity dispersion, we derive a mass contained within the half-light radius of M(r$<\rm r\textsubscript{h})=6.9^{+3.2}_{-2.8}\times10^6$~M$\textsubscript{\(\odot\)}$. This mass corresponds to a mass-to-light ratio of $\text{[M/L]}_{\rm r_{h}}=37^{+17}_{-15}$~M$_\odot$/L$_\odot$, demonstrating, for the first time, that And XXV has an unambiguous dark matter component. We also measure the metallicity of And~XXV to be $\rm[Fe/H]=-1.9\pm0.1$~dex, which is in agreement with previous results. Finally, we extend the analysis of And~XXV to include mass modelling using \texttt{GravSphere}. We find that And~XXV has a low central dark matter density, $\rho_{\rm DM}$(150 pc)= 2.7$^{+1.8}_{-1.6}\times$10$^7$~M$_\odot$ kpc$^{-3}$, making And~XXV a clear outlier when compared to other Local Group (LG) dSphs of the similar stellar mass. In a companion paper, we will explore whether some combination of dark matter cusp-core transformations and/or tides can explain And XXV’s low density.
\end{abstract}




\section{Introduction}
Dwarf galaxies are typically dark matter dominated systems, with mass-to-light ratios in the range 10-1000s, making them ideal systems for testing dark matter physics and cosmology \citep[e.g.][]{Mateo1998,Tolstoy2009,Simon2019}. The most successful cosmological model to date is Lambda Cold Dark Matter ($\Lambda$CDM), which explains the cosmic microwave background radiation \citep[e.g.][]{Peebles1982,Komatsu2009,Aubourg2015,Planck2015,Planck2020}, nucleosynthesis \citep[e.g.][]{Walker1991,Smith1993,Burles1999, Cyburt2004, Steigman2007, Fields2011} and structure formation on large scales \citep[e.g.][]{Bond1996,Springel2005,Gao2005,Gao2005b,Tegmark2006, Springel2006}, albeit by invoking three mysterious components – dark matter, dark energy and inflation \citep{Planck2015,Planck2020}. However, on galactic scales, particularly dwarf galaxies, there have been a number of long-standing tensions between $\Lambda$CDM predictions and observations \citep[e.g.][]{Bullock2017,Buckley2018, Sales2022}. One such tension is the so-called `cusp-core' problem \citep{Moore1994,Flores1994, deBlok2010, Navarro2010, Walker2011,Read2017, Genina2018, Read2019}, which arises when comparing the dark matter density profiles of dwarf galaxies from dark matter only simulations in a $\Lambda$CDM cosmology to observations of dwarf galaxies. $\Lambda$CDM predicts that the dark matter inside dwarf galaxies should follow a universal `cusped' profile that has a highly dense centre that steeply decreases with radius, such that $\rho_{\rm{DM}}\propto r^{-1}$ \citep{Dubinski1991,Navarro1996,Navarro1997,Moore1999}. Instead, a growing number of dwarfs have been observed with `cored' dark matter profiles described by a shallow central density that remains roughly constant in the centre, such that $\rho_{\rm{DM}} \sim \rm{constant}$ \citep{Flores1994,Moore1994,deBlok2001,deBlok2001b,deBlok2010,Marchesini2002,Simon2005,Battaglia2008,Walker2010,Agnello2012,Amorisco2012, Newman2013,Read2017,Read2019}. 

For many of the cored dwarfs observed, a plausible solution to the cusp-core problem within $\Lambda$CDM paradigm is dark matter heating. This is the process of sudden gas removal due to stellar winds, dynamical friction or supernovae feedback which results in gravitational fluctuations, causing the dark matter halo to expand. When this process is repeated across many cycles of star formation the dark matter halo expands irreversibly, reducing the central dark matter density and flattening the cusped profile into a core over time \citep{Navarro1996b,Gnedin2002,Read2005,Mashchenko2008,Pontzen2012,Zolotov2012,Brooks2014,Onorbe2015,Tollet2016,Read2016,Read2019}. However, dark matter heating is only proposed to be effective in galaxies with extended star formation \citep{Read2016,Read2019} that are above a stellar-mass-to-halo-mass ratio threshold of $\rm{M}_*/\rm{M_{200}}\sim5\times10^{-4}$ \citep[][but see \citet{Orkney2021}]{DiCintio2014}. 

Puzzlingly, there is a subset of dwarfs within the low surface brightness population of the LG with $\rm{M}_*/\rm{M_{200}}$ below this threshold with unusually low central densities that are unlikely to be explained by dark matter heating alone. Around the Milky Way (MW) two such dwarfs have been found, Crater~II \citep{Torrealba2016} and Antlia~II \citep{Torrealba2019}. Similarly around Andromeda (M31) two more have been observed, Andromeda~XIX \citep{McConnachie2008,Collins2020} and Andromeda~XXI \citep{Martin2009, Collins2021}. These systems have a diffuse nature, described by a large half-light radius and very low surface brightness. Furthermore, the mass contained within the half-light radius for these systems, determined from velocity dispersions measurements, is far lower than expected for their size or brightness \citep{Caldwell2017, Fu2019,Torrealba2019, Collins2020, Collins2021}, meaning that it is likely that these systems reside in low mass and low density dark matter halos. While detailed star formation histories for these systems are not currently available to completely rule out dark matter heating as the cause, even under the (unlikely) assumption of highly efficient dark matter heating, their low densities cannot be reproduced \citep{Torrealba2019,Collins2021}. Instead, tidal interactions with the host (MW or M31) are the suspected culprit causing the low density dark matter halos of these systems. 

In this paper, we investigate another potential outlier, Andromeda~XXV (And~XXV), a dSph satellite galaxy of M31 first discovered as part of the Pan-Andromeda Archaeological Survey \citep{Richardson2011}. And~XXV was previously identified as a potential outlier within the LG in a 2013 study (\citet{Collins2013}, hereafter known as \citetalias{Collins2013}) that performed a kinematic analysis of numerous M31 dSph satellites. And~XXV was found to have a low velocity dispersion, $\sigma_{\rm v}$=3.0$^{+1.2}_{-1.0}$~kms$^{-1}$, which results in a mass-to-light ratio of [M/L]$_{\rm half}$=10.3$^{+7.0}_{-6.7}$~M$_{\odot}$/L$_{\odot}$ consistent with a simple stellar system with no appreciable dark matter component, albeit with large uncertainties due to the small number of member stars (26) with stellar kinematics in their sample. As such, and as \citetalias{Collins2013} point out, more stellar velocities are needed to confirm or rule out the presence of dark matter in this system. We present a revised kinematic analysis, using the spectroscopic data set outlined in \citetalias{Collins2013} combined with new observations to provide a larger sample size, more than double that used in \citetalias{Collins2013}, allowing us to critically reassess if And~XXV really is a dSph with no appreciable dark matter. We will also extend the analysis to investigate the metallicity of And~XXV. Finally we use the dynamical mass modelling tool \texttt{BINULATOR + GravSphere} \citep{Read2017b, Read2018, Read2019, Gregory2019, Genina2020, Collins2021, Read2021}, to gain an insight into the dark matter content of And~XXV.

The outline of the paper is as follows: in $\S$\ref{sec:observations} we detail the photometric and spectroscopic observations used in this study. We detail the contaminant removal procedure used to remove MW and M31 contaminant stars from our sample in $\S$\ref{sec:contaminant removal} and discuss our kinematic analysis of And~XXV in $\S$\ref{sec:kinematic analysis}. In $\S$\ref{sec:metallicity} we investigate the metallicity of And~XXV. Next, we outline the dynamical mass modelling results for And~XXV $\S$\ref{sec:mass modeling}. In $\S$\ref{sec:discussion} we discuss our results and finally we conclude in $\S$\ref{sec:conclusions}.

\section{Observations} \label{sec:observations}
\subsection{DEIMOS spectroscopy} \label{sec:spectroscopy}
The spectroscopic data for And~XXV were obtained using the Deep Extragalactic Imaging Multi-Object Spectrograph (DEIMOS) \citet{Faber2003, Cooper2012} mounted on the Keck II telescope. The observations are comprised of two masks. The first was taken in September 2010 (previously presented in \citetalias{Collins2013}), and the second mask was observed on 18th August 2018. The instrumental setup was the same for both masks, using a 1200~line~mm$^{-1}$ grating with a resolution of 1.3~{\AA}. To determine the velocity and metallicity of each member star, we use the calcium triplet (Ca(II)) lines. The Ca(II) lines are present in the region around $\sim$8500~{\AA}. As such our observations targeted the wavelength range of $\sim$5600 - 9800~{\AA} with a central wavelength of 7800~{\AA} to resolve the Ca(II) lines. Each mask was split into 3$\times$20~minute exposures, combining to give a total exposure time of 3600~seconds per mask. The average seeing was $\sim0.8^{\prime\prime}$ for both masks, which resulted in an average signal-to-noise ratio (S/N) of $\sim$5~per pixel.

The data were reduced using a custom pipeline, described in detail in \citet{Ibata2011} and \citetalias{Collins2013}. In short, the pipeline detects and removes cosmic rays, then corrects for scattered light, slit function, illumination and fringing. To account for pixel-to-pixel variations flat-fielding was performed. Next wavelength calibrations of each pixel were conducted using arc-lamp exposures. Then, the sky was subtracted from the 2-dimensional spectra. Finally, each spectrum was extracted (without resampling) from a small spatial region around each target. The velocities and corresponding uncertainties for all stars were derived using the strong Ca(II) triplet feature found in the spectra. The non-resampled data was compared to a template Ca(II) spectrum and a most-likely velocity and uncertainty for each star were derived using a Markov Chain Monte Carlo (MCMC) routine. The final value for the velocity uncertainty comprises of the uncertainty from the MCMC posterior distribution combined with the uncertainty inherent to DEIMOS which was taken to be ${2.2}$~kms\textsuperscript{-1} \citep{SimonGeha2007}. The results from each mask were combined into a single catalogue.

 Misalignments of the slitmasks can cause velocity shifts of up to 15 km/s. We correct for these by comparing atmospheric models to the telluric absorption lines in the spectrum to shift the spectra to the correct frame as described in \citetalias{Collins2013}. To further confirm if any significant misalignment had occurred we check for a velocity error gradient across the mask. None were found in either mask. Finally, there were a few stars present both masks. The matching stars were compared to ensure the velocity uncertainties were well measured.

\subsection{LBT LBC photometry} \label{sec:photometry}
The photometric data for And~XXV were obtained using the Large Binocular Cameras (LBC) mounted on the Large Binocular Telescope (LBT) in the $V\textsubscript{Bessell}$- and $i\textsubscript{SLOAN}$-band. Observations were conducted on the nights of 28th October and 8th November, 2011 with a seeing that was ranging from 0.7" to 1.2". With its 4i say  CCDs, the cameras both cover an area of about $23\times25$ arcmin$^2$. In total, 29 exposures of 360\,s were taken with each filter, for a total exposure time of 2.9\,h per band. 

The raw photometry was reduced using the CASU pipeline where the images are debiased, flat-fielded, trimmed, and gain-corrected \citep{Irwin2001}. The reduced photometric data was made into a catalogue. Each datum was then morphologically categorised to distinguish between stellar, non-stellar (e.g. background galaxies) or noise-like objects. Only stellar objects were considered for further analysis. Finally, the data were extinction corrected using the dust maps from \citet{Schlafly2011}, using 0.271 ($V\textsubscript{Bessell}$) and 0.171 ($i\textsubscript{SLOAN}$). \\

The photometric and spectroscopic observations were combined by cross-matching the on-sky position of each star within an allowed tolerance of $\pm$1 arcsecond.

\section{Selecting members of Andromeda XXV} \label{sec:contaminant removal}
Before we can perform the kinematic analysis of And~XXV we must identify the most probable members of And XXV and remove any contaminants from the MW or M31 halo. It is difficult to differentiate between member and contaminant stars using velocity information alone. This is especially true for foreground MW stars as the two systems have similar systemic velocities. Instead, to determine likely member stars, we use a `triple-threat' probabilistic approach which was first outlined in \citet{Tollerud2012} and \citetalias{Collins2013}, then further developed in \citet{Collins2020, Collins2021} and \citet{Gregory2019}. The method assigns each star a probability of membership using three probability criteria: (1) the star's position on the sky with relation to the centre of And~XXV, $P\textsubscript{dist}$; (2) the star's position on a colour magnitude diagram (CMD) of And~XXV, $P\textsubscript{CMD}$; and (3) the velocity of the star, $P\textsubscript{vel}$. We discuss each of these criteria in more detail below.

\subsection{Distance probability}

P\textsubscript{dist} is determined using an exponential radial surface brightness profile modelled as:
\begin{equation}
    P\textsubscript{dist} = \exp{ \left[-\left(\frac{r^2 }{2{\eta\textsubscript{dist}}r\textsubscript{p}^2}\right)\right]}
	\label{eq:Pdist}
\end{equation}
where $r$ is the radial distance of the star from the centre of And XXV (taken to be 0$^{h}$ 30$^{m}$ 9.9$^{s}$ - RA, 46$^{\text{\textdegree}}$ 51$^{'}$ 41$^{"}$ - Dec \cite{Martin2016}) and $\eta\textsubscript{dist}$ is a free parameter used to scale the exponential profile to the size of And~XXV. $\eta\textsubscript{dist}$ was extensively tested for any potential biases or dependencies and a final value of $\eta\textsubscript{dist} = 2.5$ was used. $r_p$ is a Plummer profile \citep{Plummer1911} used to modify the half-light radius, $r\textsubscript{h}$, to account for any ellipticity, described by: 
\begin{equation}
    r\textsubscript{p} = \frac{r\textsubscript{h}(1-\epsilon)}{1+{\epsilon}\cos{(\theta)}}
	\label{eq:plummer profile}
\end{equation}
where $r\textsubscript{h}=2.7$~arcmin \citep{Savino2022}), $\epsilon$ is the ellipticity ($\epsilon$ = 0.03) and $\theta$ is the stars angular position with respect to the dwarfs major axis ($\theta = -16$~\textsuperscript{{\textdegree}}) \citep{Martin2016}. \\

\subsection{Colour-magnitude diagram probability}

P\textsubscript{CMD} is determined using the colour-magnitude diagram of And~XXV. A by-eye best-fit isochrone was overlaid onto the CMD to identify the stars most likely to be red giant branch (RGB) stars of And XXV. 
The isochrone used was an old, metal-poor isochrone ($\rm [Fe/H]=-1.9$~dex, $\rm [\alpha/Fe]=0.0$~dex, $\rm age=12$~Gyr) obtained from the \texttt{DARTMOUTH} stellar evolutionary models \citet{Dotter2008} shifted to the distance modulus of And~XXV, $m-M=24.38$ \citep{Savino2022}). The CMD of And~XXV and the by-eye best-fit isochrone are shown in the right panel of Figure.~\ref{fig:P_tot}.  The minimum distance, $d\textsubscript{min}$, of each star to the isochrone was converted into a probability of membership using: 
\begin{equation}
    P\textsubscript{CMD} = \exp{ \left[-\left(\frac{d\textsubscript{min}^2 }{2\eta\textsubscript{CMD}^2}\right)\right]}
	\label{eq: Piso}
\end{equation}
where $\eta\textsubscript{CMD}$ is another free parameter, this time used to account for the scatter of stars around the best-fit isochrone. Again, this free parameter was tested for any potential biases or dependencies and a final value of $\eta\textsubscript{CMD} = 0.2$ was chosen.

\subsection{Velocity probability}
P\textsubscript{vel} is determined by simultaneously fitting the velocities for all of the stars in the spectroscopic observations by assuming that these stars inhabit a profile of four dynamically distinct peaks. Four Gaussians are used to describe the different peaks which correspond to And~XXV stars (P\textsubscript{And~XXV}), then M31 halo contaminant stars (P\textsubscript{M31}) and two peaks for the MW contaminant stars (P\textsubscript{MW1}, P\textsubscript{MW2}). The MW velocity profile is often assumed to be a single distribution for contaminant removal purposes. However, due to the similarity in systemic velocity for And~XXV and the MW, for our study, this assumption is not valid. Instead, we need to include this complexity in our model by modelling the MW as two velocity distributions (e.g. \citealp{Gilbert2006}). Each Gaussian is defined by a systemic velocity ($\nu_{\rm r}$) and a velocity dispersion ($\sigma_{\rm\nu}$), such that the probability of each star belonging to each peak (P\textsubscript{And~XXV}, P\textsubscript{M31}, P\textsubscript{MW1} and P\textsubscript{MW2}) is: 
\begin{equation}
    P\textsubscript{peak} = \frac{1}{\sqrt{2\pi}\sqrt{{\sigma_{\nu_{\rm peak}}^2}+\nu\textsubscript{err,i}^2}}\times\exp{\left(-\frac{1}{2}\left[\frac{\nu\textsubscript{peak}-\nu_{i}}{\sqrt{\sigma_{\nu_{\rm peak}^2}+\nu\textsubscript{err,i}^2}}\right]^2\right)}
	\label{eq:P_peak}
\end{equation}
where $\nu\textsubscript{i}$ and $\nu\textsubscript{err,i}$ are the velocity and velocity uncertainty of a given star respectively. The overall log-likelihood function is therefore described by: 
\begin{equation}
    \log(\mathcal{L}) = \sum_{i=1}^{N}(\alpha P\textsubscript{And~XXV} + \beta P\textsubscript{M31} + \gamma P\textsubscript{MW1} + \delta P\textsubscript{MW2})
	\label{eq:likelihood}
\end{equation}
where $\alpha$, $\beta$, $\gamma$ and $\delta$ are constants describing the proportion of stars belonging to And~XXV, M31, MW1 and MW2 respectively and are normalised, such that $\alpha+\beta+\gamma+\delta=1$. The components of this likelihood function were found using $\texttt{emcee}$, a python Markov Chain Monte Carlo (MCMC) package \citep{Foreman-Mackey2013}. The routine used 200 walkers, over 5000 iterations with a burn-in stage of 1550 and uniform flat priors were introduced for each parameter, see Table.~\ref{tab:priors} for a summary of the priors and results for the MCMC analysis. 

It is important to note that, while the velocity and velocity dispersion values will likely resemble the final values, this is not the final kinematic result for And~XXV. It is only used to determine P\textsubscript{vel}. This is because it does not consider the impact of any potential M31 or MW contamination in the wings of the And~XXV Gaussian, hence why the other two probability filters are important and the combination of all three filters is used to weight the likelihood function in the final kinematic analysis. The probability distribution described by the Gaussian for each peak can then be combined to give $P\textsubscript{vel}$, using: 
\begin{equation}
    P\textsubscript{vel} = \frac{P\textsubscript{AndXXV}}{P\textsubscript{MW1}+P\textsubscript{MW2}+P\textsubscript{M31}+P\textsubscript{AndXXV}}
	\label{eq: Pvel}
\end{equation}
\\

\begin{table}
	\centering
	\caption{Prior values and results for the variables used in our \texttt{emcee} analysis for P\textsubscript{vel}. \text{[*]}\textit{ Note: This is not the final systemic velocity and velocity dispersion value for And~XXV - it is only used to determine P\textsubscript{vel}}}
	\label{tab:priors}
	\begin{tabular}{lll}
		\hline
		\textbf{Priors:} & & \\
		Peak & $\nu_{\rm r}$ (kms$^{-1}$) & $\sigma_{\rm\nu}$ (kms$^{-1}$) \\
		\hline
		P\textsubscript{And~XXV} & $-130<\nu_{\rm r}<-90$  & $0<\sigma_{\rm\nu}<50$ \\
		P\textsubscript{M31} & $-400<\nu_{\rm r}<-130$ & $0<\sigma_{\rm\nu}<200$\\
		P\textsubscript{MW1} & $-60<\nu_{\rm r}<50$ & $0<\sigma_{\rm\nu}<100$\\
		P\textsubscript{MW2} & $-90<\nu_{\rm r}<-50$ & $0<\sigma_{\rm\nu}<100$\\
		\hline
		\hline
		\textbf{Results:} & & \\
		Peak & $\nu_{\rm r}$ (kms$^{-1}$) & $\sigma_{\rm\nu}$ (kms$^{-1}$)\\
		\hline
		\vspace{0.12cm}
		P\textsubscript{And~XXV} & $-107.6\pm{1.4}$  & $5.9^{+2.1}_{-1.7}$ [*]\\
		\vspace{0.12cm}
		P\textsubscript{M31} & $-268.2^{+49.5}_{-48.0}$ & $132.6^{+29.8}_{-27.0}$\\
		\vspace{0.12cm}
		P\textsubscript{MW1} & $-45.7^{+15.5}_{-9.3}$ & $28.3^{+13.2}_{-9.8}$\\
		\vspace{0.12cm}
		P\textsubscript{MW2} & $-66.7^{+10.8}_{-14.2}$ & $36.9^{+16.0}_{-8.3}$\\
		 
	\end{tabular}
\end{table}

\begin{figure*}
	\includegraphics[width=\textwidth]{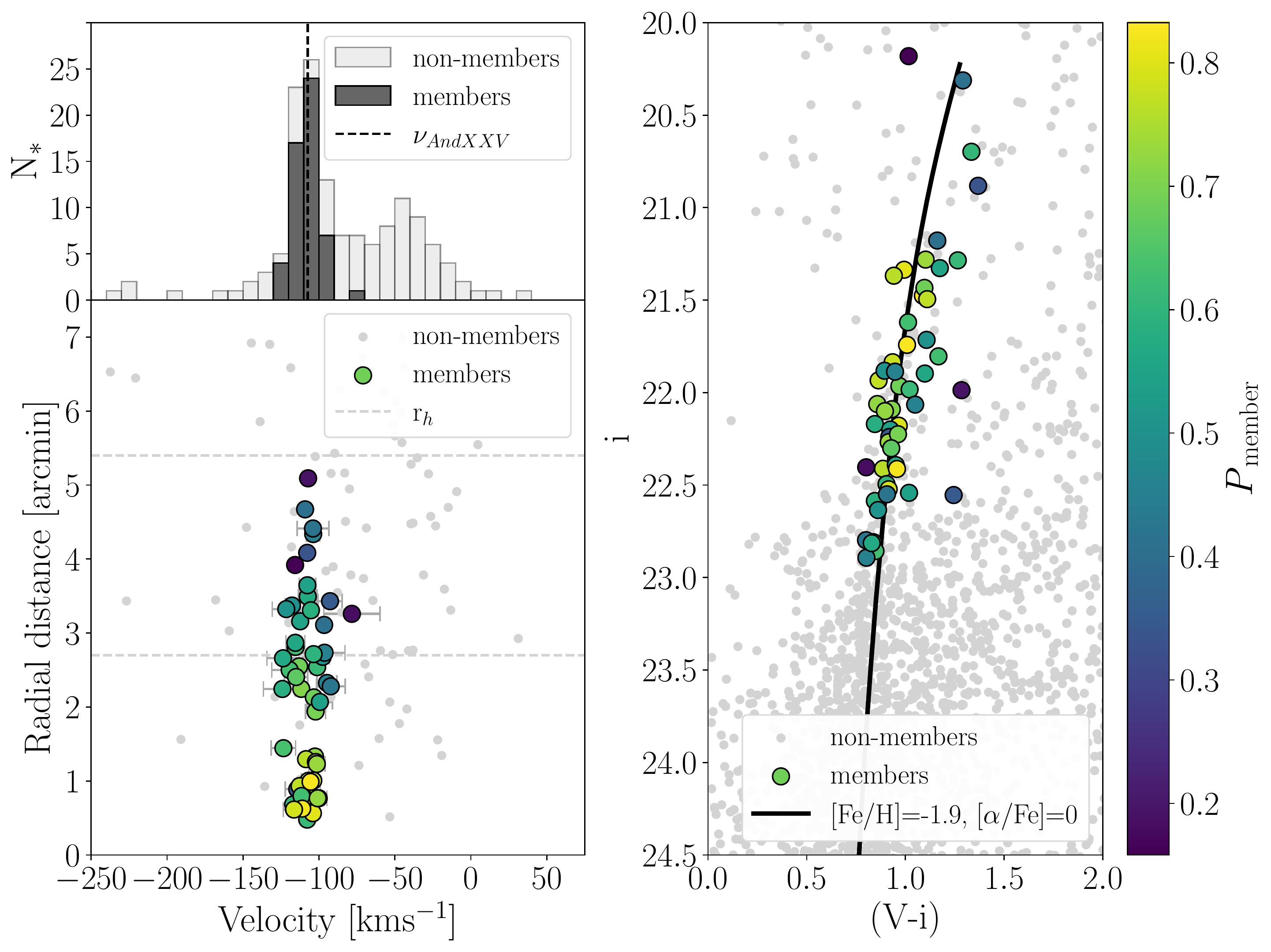}
      \caption{{\bf Left Top}: The velocity histogram from And~XXV. Light grey represents non-members likely contamination from foreground MW stars or M31 halo stars. Dark grey represents the 53 identified member stars of And~XXV. The black dashed line represents the systemic velocity of And~XXV determined in this study. {\bf Left Bottom:} Radial distance from the centre of And~XXV as a function of the line of sight velocity. The smaller light grey circles are the non-member stars from the spectroscopic data. The larger coloured circles are the member stars, colour-coded by the probability of membership (see colour bar on the right) and the error bars are the 1$\sigma$ uncertainties. The horizontal grey dashed lines represent 1$\times$ and 2$\times$ r$_h$ from the bottom up respectively. {\bf Right: } The colour magnitude diagram for And~XXV. Again the smaller light grey circles are non-member stars within 1.5$\times$ r$_h$ and the larger coloured circles are the member stars colour-coded by the probability of membership. The black solid line is a by-eye best-fit isochrone ($\rm [Fe/H]=-1.9$~dex, $\rm [\alpha/Fe]=0.0$~dex, $\rm age=12$~Gyr) for And~XXV taken from the \texttt{DARTMOUTH} stellar evolutionary models \citet{Dotter2008} and shifted to the distance modulus of And~XXV, $m-M=24.38$ \citep{Savino2022}).}
    \label{fig:P_tot}
\end{figure*}

The final probability of membership to And~XXV for each star is the product of the three probability cuts, such that:
\begin{equation}
    P\textsubscript{member} = P\textsubscript{dist} \times P\textsubscript{CMD} \times P\textsubscript{vel}
	\label{eq: Probabilty of Membership}
\end{equation}
Stars with a probability of $P\textsubscript{member} > 0.15$ were considered member stars. The probability of membership cut-off is kept intentionally low as we use the velocity dispersion for the kinematic analysis. Hence, we do not want to artificially decrease the velocity dispersion by removing potential candidates with a probability cut that is too strict. Furthermore, all non-members have a probability of membership significantly below this cut-off point. A total of 53 members were identified (as shown in Fig.~\ref{fig:P_tot}). This data contains more than double the number of members compared to that used in \citetalias{Collins2013}.

\section{Kinematic Analysis of Andromeda XXV} \label{sec:kinematic analysis}
We used another $\texttt{emcee}$ routine to determine the velocity ($\nu$) and velocity dispersion ($\sigma_{\rm\nu}$) of And~XXV. This time, the routine fits a single Gaussian for the 53 member stars, with the likelihood of each star weighted by the star's respective probability, such that the log-likelihood function is:
\begin{equation}
    \log(\mathcal{L}) = \sum_{i=1}^{N}\log(P\textsubscript{{member}\textsubscript{i}} P\textsubscript{{AndXXV}\textsubscript{i}})
	\label{eq: final gauss}
\end{equation}

\begin{figure}
	\includegraphics[width=\columnwidth]{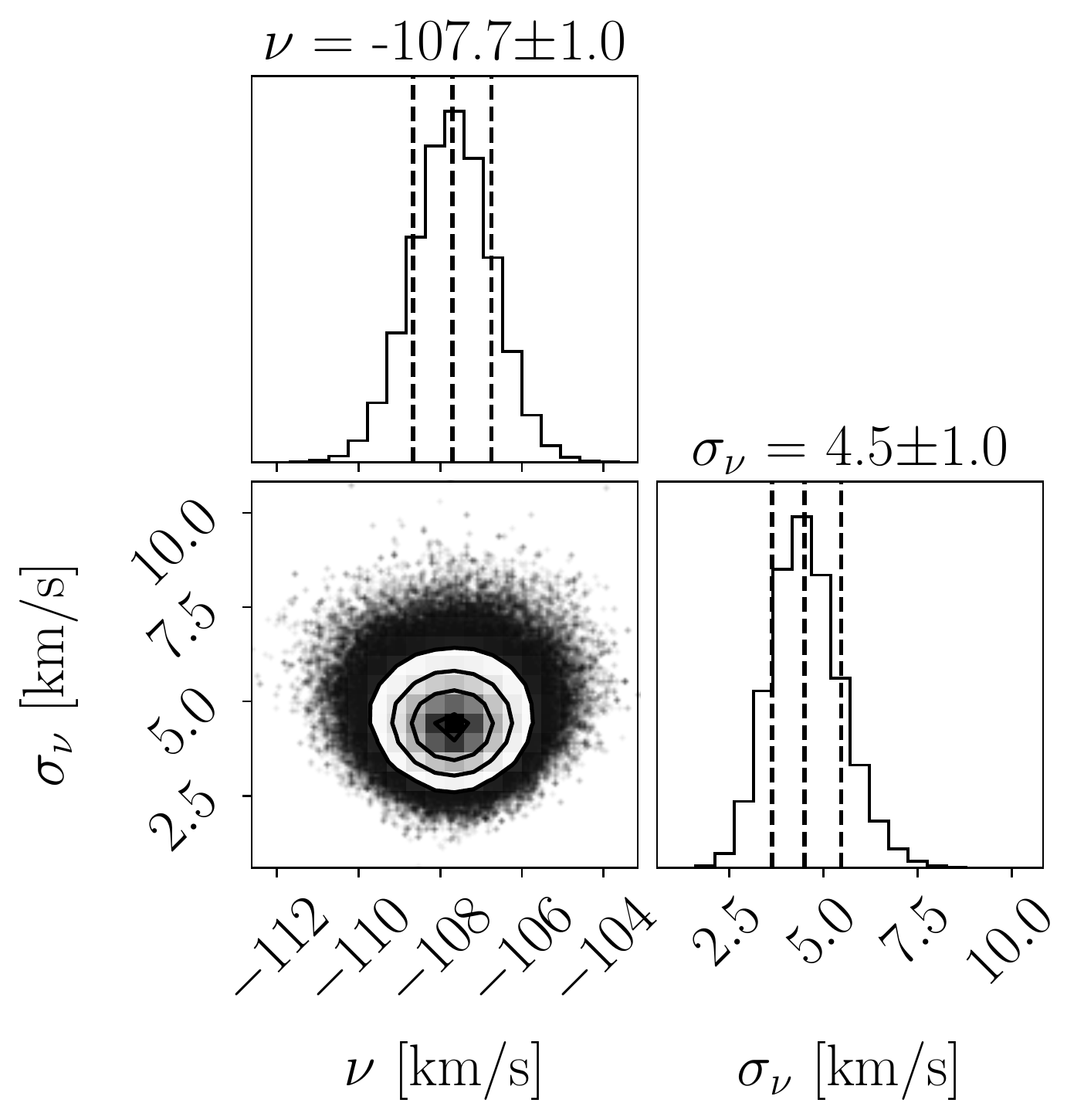}
      \caption{A corner plot showing the results of our kinematic analysis using \texttt{EMCEE}. The velocity and velocity dispersion are well resolved, giving $\nu_r=-107.7\pm1.0$~km~s$^{-1}$ and $\sigma_{\nu}=4.5\pm1.0$~kms$^{-1}$}
    \label{fig:corner}
\end{figure}

The routine used 500 walkers over 5000 iterations with a burn-in stage of 3750. The values of $\nu$ and $\sigma_{\rm\nu}$ for And~XXV from the $P\textsubscript{vel}$ analysis were used as initial guesses and the same flat priors used for $P\textsubscript{vel}$ were implemented. The resulting posterior distribution can be seen in Fig.~\ref{fig:corner}.

\subsection{Kinematic results} \label{sec:kinematic results}

The velocity and velocity dispersion are well resolved, giving $\nu=-107.7\pm{1.0}$~km~s$^{-1}$ and $\sigma_{\rm\nu}=4.5\pm{1.0}$~km~s$^{-1}$, the uncertainties are
the 1$\sigma$ uncertainty intervals from the posterior distributions. The properties derived for And~XXV are listed in Table~\ref{tab:AndXXv properties}. The velocity and velocity dispersion are consistent with those derived in \citetalias{Collins2013} ($\nu=-107.8\pm{1.0}$~km~s$^{-1}$, $\sigma_{\rm\nu}=3.0^{+1.2}_{-1.0}$~km~s$^{-1}$) although our velocity dispersion is higher. This is likely due to the targeting of the \citetalias{Collins2013} study. The first mask preferentially targeted stars close to the centre of And XXV. As seen in Fig.~\ref{fig:P_tot}, the central stars are kinematically colder than those in the outskirts (velocity dispersion of $\sim$3.3~km~s$^{-1}$ in the centre vs $\sim$5.7~km~s$^{-1}$ on the outskirts). The outermost stars included in the new spectroscopic data set increase the average dispersion measured in this study. It is important to note that at the outskirts of And~XXV we are more susceptible to contamination from non-member stars, especially due to the similarity in velocity between And~XXV and the foreground MW stars. We extensively tested out membership selection, however, it is possible inclusion of contaminates could potentially contribute to the velocity flaring at larger radii.

\begin{table}
	\centering
	\caption{The properties of And~XXV. Sources a: \citet{Martin2016}, b: \citet{Savino2022}, c: this work.}
	\label{tab:AndXXv properties}
	\begin{tabular}{lll}
		\hline
		Property &Value &Source \\
		\hline
		$\rm\alpha$,$\rm\delta$ (J2000) & 00:30:09.9 +46:51:41 & a\\
		\vspace{0.1cm}
		m$_V$ & 15.3$^{+0.3}_{-0.2}$ & a\\
		\vspace{0.1cm}
		M$_V$ & -9.1$^{+0.3}_{-0.2}$ &b \\
		\vspace{0.1cm}
		D (kpc) & 751.6$^{+25}_{-21}$ &b \\
		\vspace{0.1cm}
		r$_{\rm h}$ (arcmin) & 2.7$^{+0.4}_{-0.3}$&b \\
		\vspace{0.1cm}
		r$_{\rm h}$ (pc) & 590$^{+90}_{-47}$ &b \\
		\vspace{0.1cm}
		L (L$_\odot$) &3.7$^{+1.4}_{-0.5}\times$10$^5$ &b\\
		\hline
		$\nu$ (kms$^{-1}$) & -107.7$\pm$1.0 & c \\
		$\sigma_\nu$ (kms$^{-1}$) & 4.5$\pm$1.0 & c \\
		\vspace{0.1cm}
		M(r<r$_{\rm h}$) (M$_\odot$) & 6.9$^{+3.2}_{-2.8}\times$10$^6$& c \\
		\vspace{0.1cm}
		\text{[M/L]}$_{\rm r_{h}}$ (M$_\odot$/L$_\odot$) &37$^{+17}_{-15}$ & c \\
		\vspace{0.1cm}
		\text{[Fe/H]} (dex) & -1.9$\pm0.1$ & c \\
		\vspace{0.1cm}
		$\rho_{\rm DM}$(150 pc) (M$_\odot$ kpc$^{-3}$) & 2.7$^{+1.8}_{-1.6}\times$10$^7$& c \\
		\hline
	\end{tabular}
\end{table}

Using the now better constrained velocity dispersion, we can measure the mass, M($<r_\textsubscript{h}$), and mass-to-light ratio, $\text{[M/L]}_{\rm r_{h}}$, contained within the half-light radius of And~XXV. Assuming a flat velocity dispersion profile, $M\textsubscript{r\textsubscript{h}}$ can be calculated using \cite{Walker2009}:
\begin{equation}
    M\textsubscript{r\textsubscript{h}} = 580\,r\textsubscript{h}\,\sigma_{\nu}^2
	\label{eq:r_h mass}
\end{equation}

The mass of And XXV was determined to be $M\textsubscript{r\textsubscript{h}} = 6.9^{+3.2}_{-2.9}\times10^6$~M$\textsubscript{\(\odot\)}$, which is in agreement with the \citetalias{Collins2013} study within the 1$\sigma$ uncertainties. Using the updated absolute magnitude value of And~XXV, $M\textsubscript{V}=-9.1$ \citep{Savino2022}, which is equal to a luminosity of $L=3.7 ^{+1.4}_{-0.5}\times10^5$~L$\textsubscript{\(\odot\)}$, the mass-to-light ratio contained within the half-light radius is $[M/L]\textsubscript{r\textsubscript{h}}=37^{+17}_{-15}$~M$\textsubscript{\(\odot\)}$/L$\textsubscript{\(\odot\)}$. The mass-to-light ratio agrees with the value from \citetalias{Collins2013} within the 2$\sigma$ uncertainties. If we instead use the previous luminosity value ($L=6.8\times10^5$~L$\textsubscript{\(\odot\)}$) with the updated velocity dispersion we derive a mass-to-light ratio of $[M/L]\textsubscript{r\textsubscript{h}}=21^{+10}_{-9}$~M$\textsubscript{\(\odot\)}$/L$\textsubscript{\(\odot\)}$, which is consistent within 1$\sigma$. This demonstrates that the moderate increase in the mass-to-light ratio is predominantly due to the updated luminosity value that is almost half the previous value that was used in \citetalias{Collins2013}. 

\subsection{Comparison to Kinematics of Local Group dSphs}  \label{sec:kinematic LG}

\begin{figure*}
	\includegraphics[width=\columnwidth]{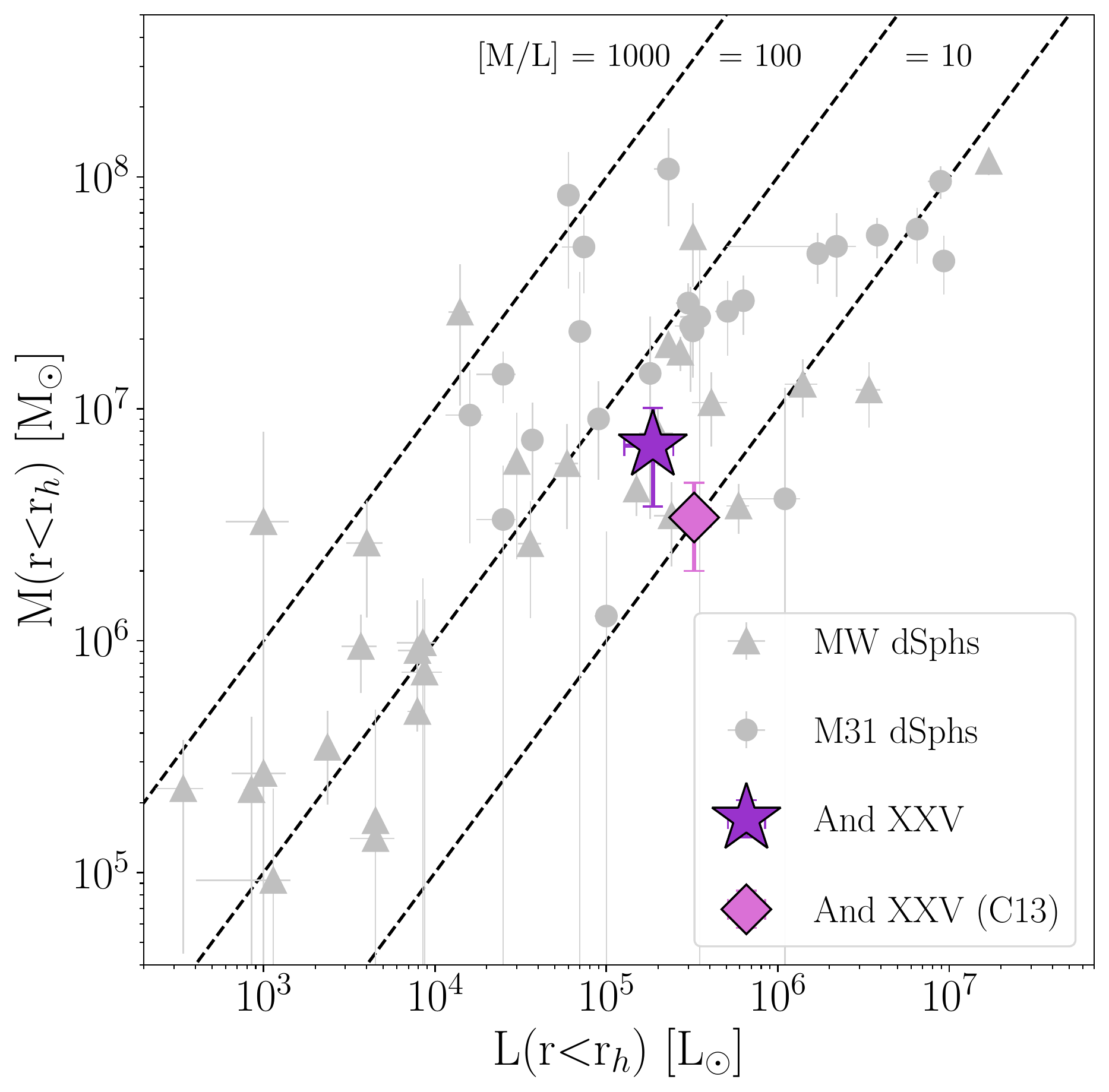}
	\includegraphics[width=\columnwidth]{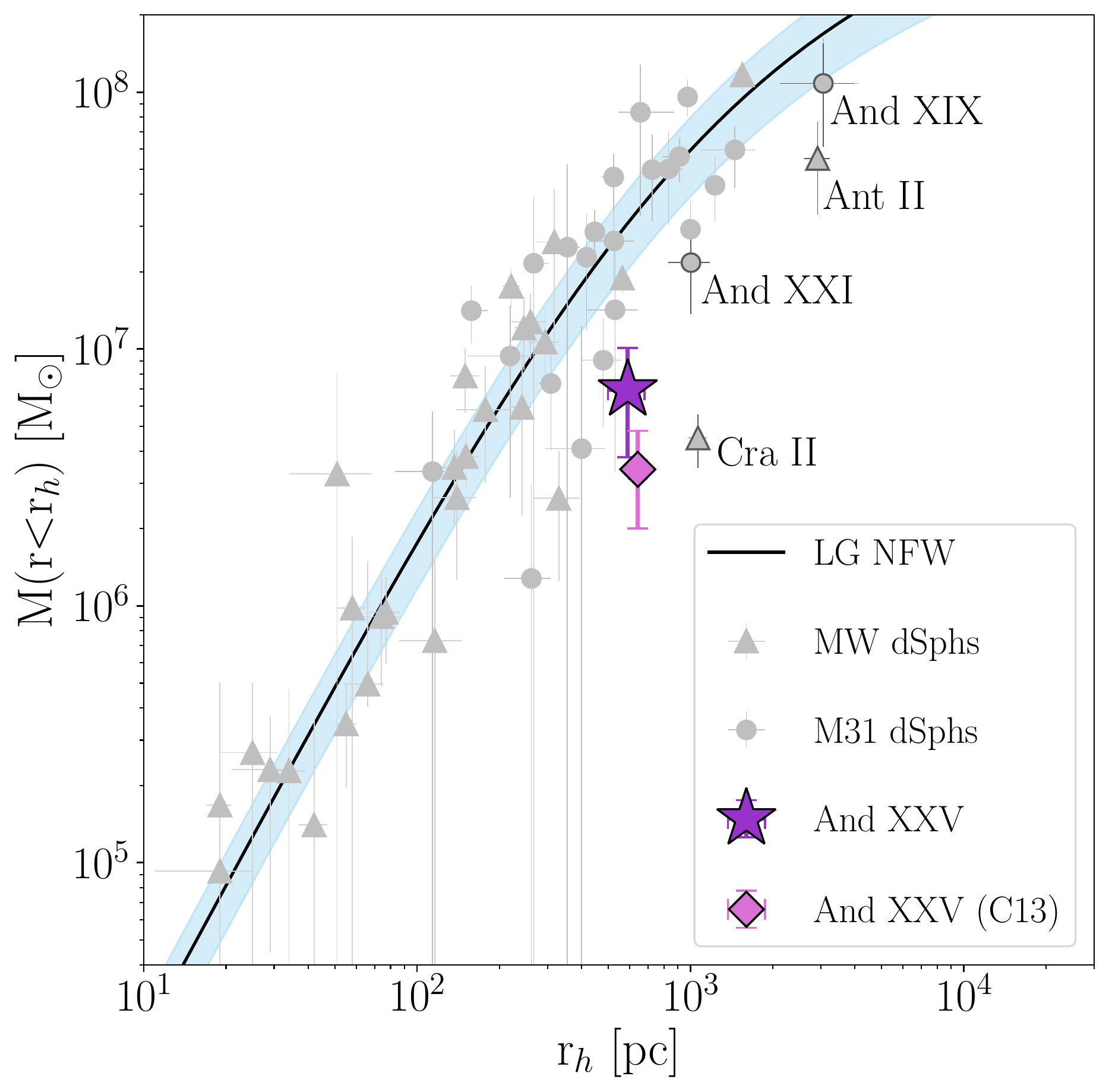}
    \caption{{\bf Left:} Mass-to-light ratios of dSphs in the LG. The diagonal black lines illustrate a mass-to-light ratio of 10, 100 and 1000 from right to left respectively. {\bf Right: } The mass contained within the half-light radius for the LG dSphs. The solid black line is the best-fitting NFW mass profile for the LG population \citep{Collins2014}, with the 1$\sigma$ uncertainties for this relationship shown by the blue shaded region. {\bf Both: }The light grey triangles are MW dSphs and the light grey circles are M31 dSphs. The \citetalias{Collins2013} result is illustrated by the pink diamond and the updated value from this study for And XXV is shown by the purple star. For all data points, the error bars show the 1$\sigma$ uncertainties.}
    \label{fig:MLMrh}
\end{figure*}

Dwarf galaxies are typically dark matter dominated systems at all radii and as such we would expect them to have mass-to-light ratios of $[M/L]\textsubscript{r\textsubscript{h}}>10$~M$\textsubscript{\(\odot\)}$/L$\textsubscript{\(\odot\)}$. The left panel of Fig.~\ref{fig:MLMrh} shows the mass-to-light ratios of LG dSphs, the light grey circles and triangles are M31 and MW dSphs respectively, taken from \cite{Tollerud2012,Collins2013,Collins2017,Collins2020,Collins2021,Walker2007,Walker2009,SimonGeha2007,Simon2011,Simon2015,Martin2007,Martin2013,Martin2013b,Martin2014,Ho2012,Kirby2015,Kirby2017}. The pink diamond shows the \citetalias{Collins2013} result. The mass-to-light ratio is indicative of a simple stellar system with no appreciable dark matter content. The purple star shows the mass-to-light ratio of And~XXV derived from this study. We can see that with the combination of the slightly increased velocity dispersion and updated luminosity used in this study, the mass-to-light ratio of And~XXV moves in line with other LG dSphs. We have shown for the first time that And~XXV has an unambiguous dark matter component and is indicative of a dark matter dominated dwarf galaxy. 

 The right panel of Fig.~\ref{fig:MLMrh} shows the mass contained within the half-light radius as a function of the half-light radius for the LG dSphs. Again the light grey circles and triangles are M31 and MW dSphs respectively, taken from the same sources. With the updated velocity dispersion the half-light radius mass moves more in line with what we would expect (compare the pink diamond - \citetalias{Collins2013} to the purple star -this study). Despite this, we can see that the mass of And~XXV is still significantly lower when compared to the best-fitting NFW mass profile for the LG population \citep{Collins2014}, illustrated by the solid black line and the light blue shaded region is the 1$\sigma$ uncertainty. This extended radial behaviour with significantly less mass than expected for its size is also shown by the other four outliers And~XXI \citep{Collins2021}, And~XIX \citep{Collins2020}, Crater~II \citep{Torrealba2016} and Antlia~II \citep{Torrealba2019}, as highlighted in Fig.~\ref{fig:MLMrh}. For these systems, it is suspected that tides have caused the low masses. Furthermore, \citetalias{Collins2013} noted that tidal interactions could have acted to lower the central density of And XXV, and the increase in velocity dispersion with radius could indicate a tidal influence. 

To investigate if we can observe the impact of tides in the kinematics, we altered the Gaussian component of the likelihood function described in Equ.~\ref{eq:P_peak} to include a velocity gradient component that would indicate the presence of tidal streams, following the methodology outlined in \citet{Martin2010}. No statistically significant velocity gradient was found. It is important to note that this does not mean And~XXV has not experienced tidal influences, especially as we potentially run into the same small number statistics issues reported by the \citetalias{Collins2013} as we are trying to fit more parameters than the simple Gaussian outlined in Equ.~\ref{eq:P_peak}. We return to the potential impact of tides on And~XXV in $\S$~\ref{sec:discussion}. 

\section{Metallicity of Andromeda XXV} \label{sec:metallicity}
We determined the metallicity of And XXV by measuring the equivalent widths of the calcium II triplet lines from the spectroscopic observations. These absorption lines are a good proxy for iron abundance [Fe/H] \citep{Armandroff1991}. We only include member stars with a good signal-to-noise ratio (S/N). For our data set, this was determined to be $S/N > 5$ per pixel, which resulted in a sample that comprised of 41 stars. To calculate the equivalent widths, we followed the methodology outlined in \citetalias{Collins2013}. Firstly, we apply a Doppler correction to each star to ensure the spectra are measured in the rest frame. The spectra were then interpolated onto a common framework before co-adding the S/N weighted spectra. Next, the spectra were normalised, such that the mean continuum was equal to one. Finally, we simultaneously fit the continuum and three Gaussian peaks to each of the three Ca II triplet lines to obtain the equivalent widths. 


\begin{figure}
	\includegraphics[width=\columnwidth]{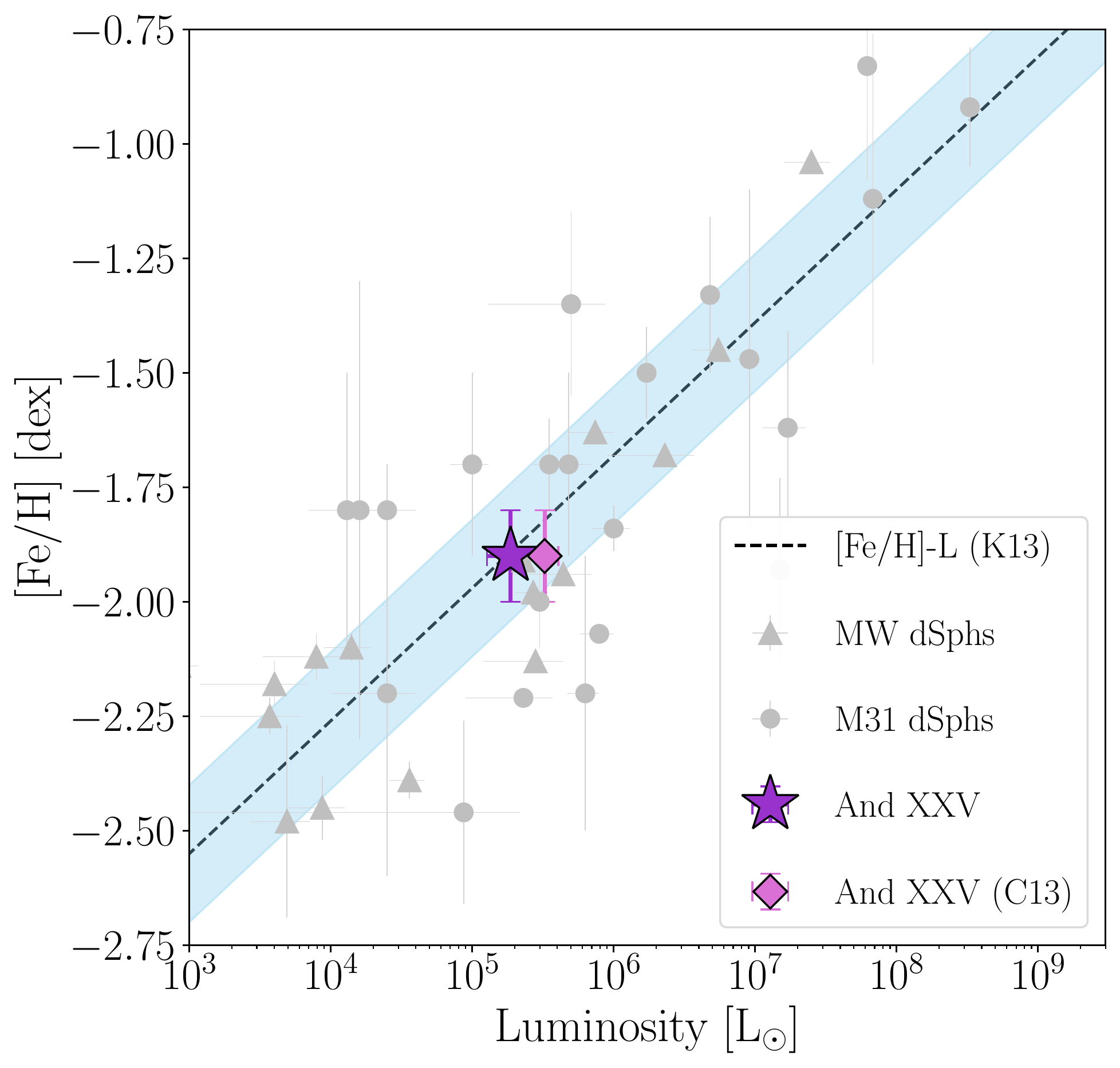}
      \caption{The luminosity-metallicity relation for LG dSphs. MW dSphs are represented by light grey triangles whereas the M31 dSphs are the light grey circles. The pink diamond shows the results for And~XXV from the \citetalias{Collins2013} study and the purple star shows the result for And~XXV from this study derived from the coadd spectra. The black dash line is the best fit luminosity-metallicity relation from \citet{Kirby2013} with the light blue band illustrating the 1$\sigma$ scatter. And~XXV is in perfect agreement with this relation. }
    \label{fig:metal comp}
\end{figure}

The metallicity was determined from the equivalent widths using the approach described by \citet{Starkenburg2010} such that:
\begin{equation}
   \text{[Fe/H]} = -2.87 + 0.195~M + 0.48~\Sigma Ca - 0.913~\Sigma Ca^{-1.5} + 0.00155~\Sigma{Ca}~M
	\label{FeH}
\end{equation}
where $\Sigma \rm Ca=$ $0.5\text{EW}_{8498}+\text{EW}_{8542}+0.6\text{EW}_{8662}$ and $M$ is the absolute magnitude of each star given by:
\begin{equation}
   M = V - 5\times\log_{10}(D_{\odot})+5
	\label{FeH_mag}
\end{equation}
where $V$ is the V-band magnitude of the star and $D_{\odot}$ is the heliocentric distance of the star which, for all stars, was assumed to be the heliocentric distance of And XXV, 751.6~kpc \citep{Savino2022}. This resulted in a metallicity for And XXV of $\rm[Fe/H]=-1.9\pm{0.1}$~dex. This value is in direct agreement with the result from the \citetalias{Collins2013} study. Furthermore, this result is in agreement with the luminosity-metallicity relation \citep{Kirby2013} for LG dwarfs, as shown in Fig.~\ref{fig:metal comp} with data taken from the sources mentioned for Fig.~\ref{fig:MLMrh} in addition to \citet{Ho2015} and \citet{Wojno2020}.

\begin{figure}
	\includegraphics[width=\columnwidth]{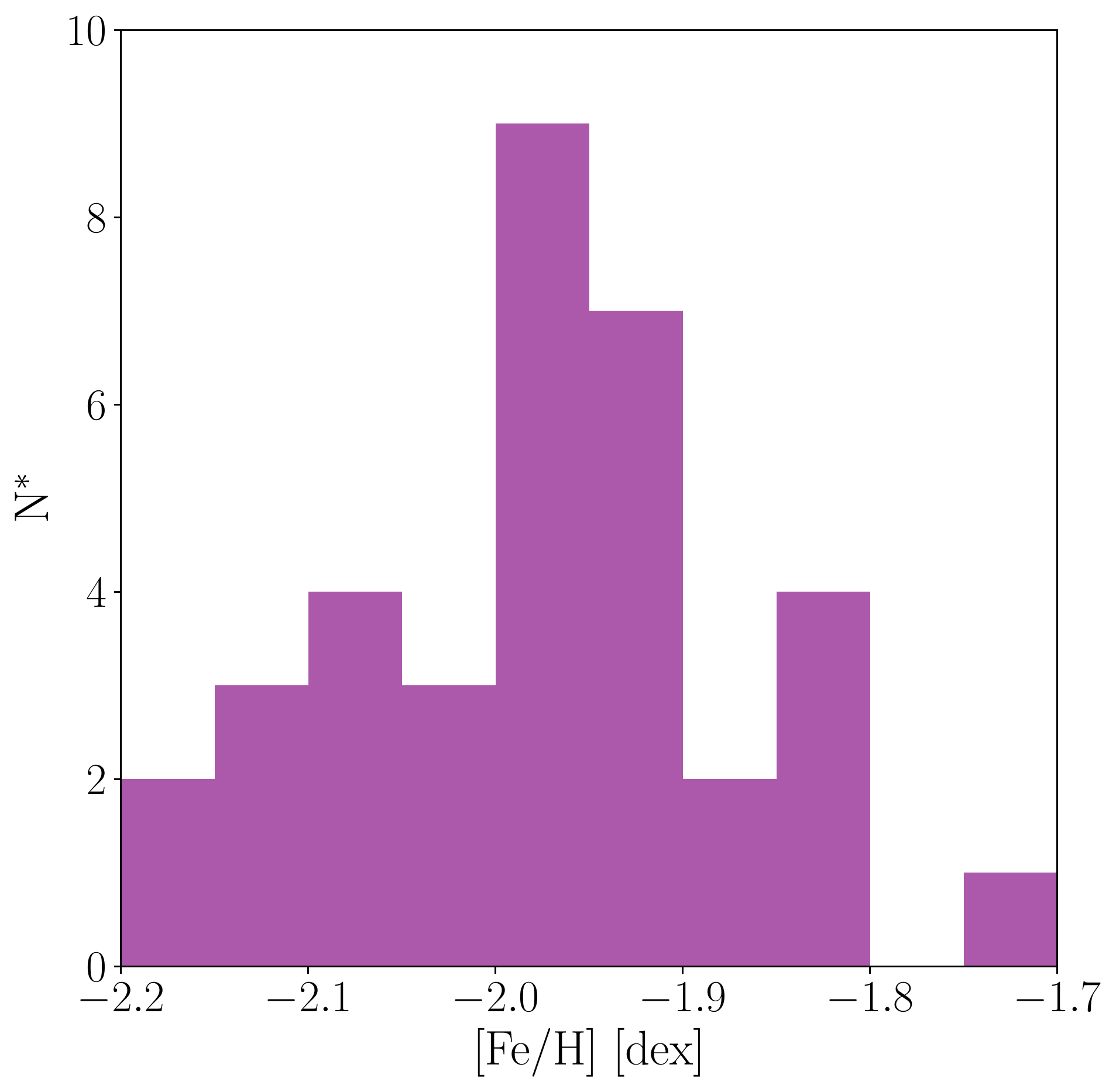}
	\includegraphics[width=\columnwidth]{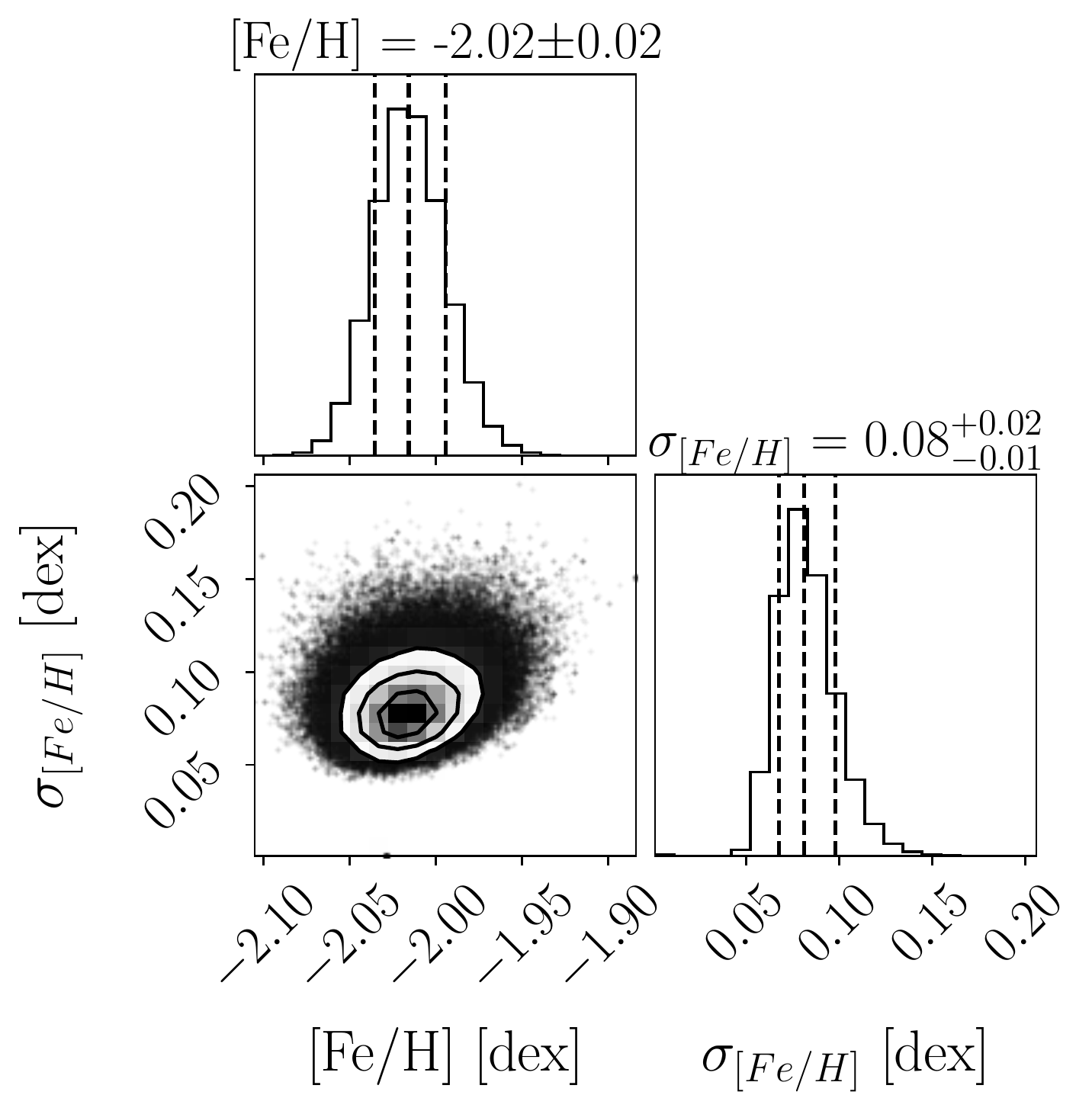}
      \caption{{\bf Top:} The metallicity distribution function derived from the individual spectra of And~XXV member stars with a S/N > 5~per pixel. {\bf Bottom:} A corner plot showing the mean and spread of metallicity for And~XXV. We find a mean metallicity of $\rm{[Fe/H]}=-2.02 \pm 0.02$~dex and a spread of $\sigma_{\rm{[Fe/H]}} = 0.08^{+0.02}_{-0.01}$~dex.}
    \label{fig:metalMDF+corner}
\end{figure}

With high S/N spectra, it is possible to obtain metallicities for individual stars. To do this we performed the same fitting procedure described above but this time for each of the 41 spectra separately. The fitting procedure was successful for 36 of the 41 spectra. The top panel in Fig.~\ref{fig:metalMDF+corner} shows the metallicity distribution function (MDF) for And XXV. An MCMC routine fitting a Gaussian to the MDF found the mean metallicity to be $\rm[Fe/H]=-2.02\pm0.02$~dex and the metallicity dispersion to be $\sigma\textsubscript{[Fe/H]} = 0.08^{+0.02}_{-0.01}$~dex, see the corner plot Fig.~\ref{fig:metalMDF+corner}. The values obtained from the MDF are also in agreement with the results from the coadded spectra, the \citetalias{Collins2013} study and the luminosity-metallicity relation \citep{Kirby2013}.

\section{Mass Modelling of Andromeda XXV}  \label{sec:mass modeling}
\subsection{\texttt{BINULATOR} + \texttt{GravSphere} - mass modeling tool} \label{sec:gravsphere}
\texttt{GravSphere}\footnote{The updated \texttt{GravSphere} code along with the new BINULATOR binning method, described in this paper and \citet{Collins2021}, is
available to download from \url{https://github.com/justinread/gravsphere}. \texttt{pyGravSphere}, a free form mass version of the \texttt{GravSphere} code \citep{Genina2020}, is available to download from \url{https://github.com/AnnaGenina/pygravsphere}.} is the dynamical mass modelling tool we used to measure the dark matter density profile of And XXV and is described described in detail in \citet{Read2017b, Read2018, Read2021, Genina2020, Collins2021}. In this section, we briefly review the \texttt{GravSphere} methodology and describe its application to And XXV. \texttt{GravSphere} solves the projected spherical Jeans equations \citep{Jeans1922,Binney1982} for a set of tracers, in this instance the 53 member stars of And~XXV, to determine the dark matter density profile, assuming that it is a spherical, non-rotating system. These equations are known to have a mass-velocity anisotropy degeneracy for which a wide range of solution combinations can satisfy the model \citep{Merrifield1990,Wilkison2002,Lokas2003,deLorenzi2009}. Several methods have been proposed to break this degeneracy (see e.g. \citet{Read2017b}. \texttt{GravSphere} addresses it by fitting two higher order “Virial Shape Parameters” (VSPs), first proposed by \citet[][see also \citet{Richardson2014}]{Merrifield1990}. The velocity anisotropy profile is difficult to constrain observationally. However, using VSPs we only need line of sight velocities (easily obtained from spectroscopic observations) to place meaningful constraints on the velocity anisotropy profile, hence breaking the mass-velocity anisotropy degeneracy. \texttt{GravSphere} uses a symmetric version of the velocity anisotropy profile, $\tilde\beta$, is used to avoid issues with infinite values \citep{Read2006}, where $\tilde\beta = 0$ describes an isotropic velocity distribution, $\tilde\beta = -1$ a complete tangential and $\tilde\beta = 1$ a complete radial distribution. Finally, in line with previous studies \citep{Read2018, Collins2021}, we adopt the \texttt{CORENFWTIDES} model to describe the dark matter distribution. This mass profile includes within its parameterisation the cusped Navarro-Frenk-White profile (NFW; Navarro et al. 1996). The NFW profile gives a good fit to dark matter density profiles in pure dark matter simulations and is described by two parameters: a virial mass $M_{200}$ and concentration parameter $c_{200}$. The \texttt{CORENFWTIDES} model adds four new parameters to this: $n$, $r_c$, $r_t$ and $\delta$. The first two control how “cored” or “cusped” the dark matter profile is inside $r_c$, where $n=1$ corresponds to a constant density flat core, $n=0$ corresponds to an $r^{-1}$ cusp, as in the NFW profile, and $n=-1$ corresponds to an even steeper $r^{-2}$ cusp. The second two parameters model the effect of tidal forces from a larger host galaxy stripping some of the outer mass away, where $r_t$ is the tidal radius beyond which the density falls off as $r^{-\delta}$.

\texttt{GravSphere} has been rigorously tested on a wide variety of mock data \citep{Read2017b,Read2018,Read2021,Genina2020,Collins2021} and stands up well when compared to other dynamical mass modeling tools \citep{Read2021}. However, for systems with a small number of stars, and/or where the velocity uncertainty is large, the binning method in the previous version of \texttt{GravSphere} can become slightly biased towards cusped profiles \citep{Gregory2019,Zoutendijk2021,Collins2021}. To resolve this issue \texttt{GravSphere} was updated to include a new separate binning routing, \texttt{BINULATOR}, first introduced and outlined in detail in \citet{Collins2021}. This update reduces the aforementioned biases by fitting a generalised Gaussian probability distribution to each bin, providing a robust estimate of the mean, variance, kurtosis and corresponding uncertainties for each bin, even for systems with an extremely small sample size - such as And~XXV. These estimates are then used as input to \texttt{GravSphere}. Detailed tests of this updated version of \texttt{BINULATOR}+\texttt{GravSphere} can be found in \citet{Collins2021} Appendix.~A.

\subsection{Implementing \texttt{BINULATOR} + \texttt{GravSphere} for Andromeda XXV} \label{sec:core and tides}
The surface brightness profile for And XXV was constructed from photometric data obtained through the LBT imaging described in $\S$\ref{sec:photometry}. We included point sources from the entire data set out to $\sim5\times r_h$. For each star the radial distance from the centre of And~XXV was measured and a probability of membership was assigned using the $P_\text{{dist}}$ and $P_\text{{iso}}$ method outlined in Equ.~\ref{eq:Pdist} and Equ.~\ref{eq: Piso}, albeit with a looser probability constraint (P$_{\text{member}}>0.01$) and larger $\eta$ parameters  to ensure a representative and complete surface brightness profile. The velocity dispersion profile was constructed from the spectroscopic data for the 53 identified members of And~XXV, using the velocity, velocity uncertainty, radial distance from the centre and the probability of membership determined using Equ.~\ref{eq: Probabilty of Membership} for each member star. For both data sets the probability of membership is summed to give the total number of `effective' tracers:  
\begin{equation}
  N\textsubscript{eff} = \sum^{N\textsubscript{mem}}_{i=0}P\textsubscript{mem,i}
	\label{eff tracer}
\end{equation}
giving $N\textsubscript{eff} = 2450$ for the photometric data and $N\textsubscript{eff} = 32$ for the velocity data, which were then split into 98 bins of 25 and 4 bins of 8 respectively. \texttt{GravSphere} fits the surface brightness profile and radial velocity profile from the kinematic and photometric data input using an \texttt{emcee} routine (see Appendix~\ref{sec:appendix}). For the parameters in the \texttt{CORENFWTIDES} profile we implemented the following priors: $7.5<\log_{10}(M_{200}/\text{M}_{\odot})<11.5$, $7<c_{200}<53$, $-2<\log_{10}(r_{c}/\text{kpc})< 10$,  $1<\log_{10}(r_{t}/\text{kpc})< 20$, $3<\delta< 5$ and $-1<n<1$ . For the symmetric velocity anisotropy, $\tilde\beta$, the priors were: $-0.1<\tilde\beta_{\infty}< 1$, $-2<\log_{10}(r_{0}/\text{kpc})< 0$ and $1<q<3$. Finally, for the stellar mass of And~XXV we convert the updated luminosity value from \cite{Savino2022} using the assumption of a stellar mass-to-light ratio of 1, in line with \citet{McConnachie2012}. This gives a stellar mass of $M_*=3.7^{+0.3}_{-0.4} \times 10^5$~M$_{\odot}$.

\begin{figure}
	\includegraphics[width=\columnwidth]{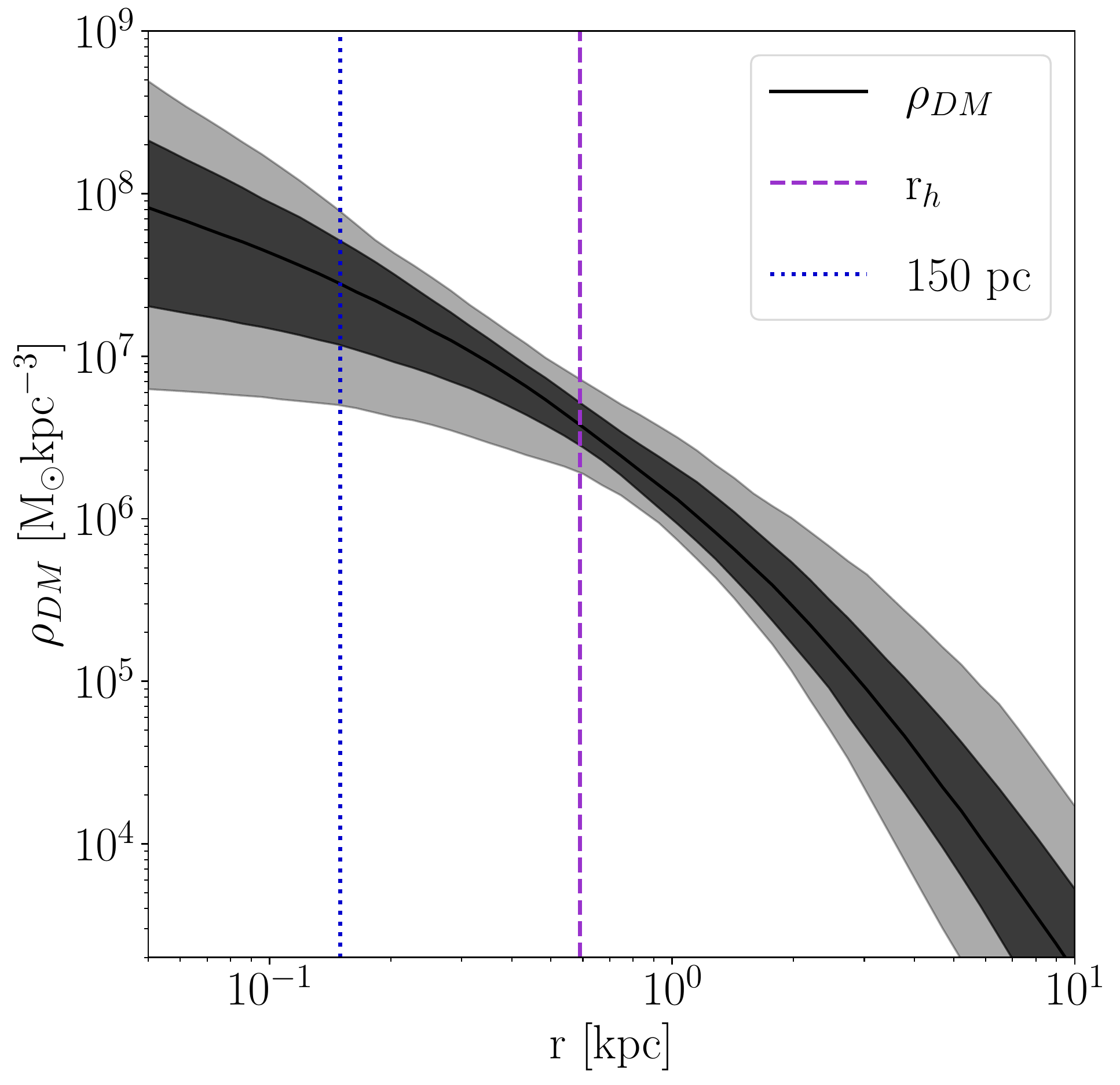}
    \caption{The dark matter density profile for And~XXV. The dark grey and light grey shaded regions are the 1$\times$ and 2$\times~\sigma$ uncertainties respectively and the purple dashed line is the literature half-light radius of And~XXV \citep{Savino2022}. The blue dotted line corresponds to 150~pc, the radial distance at which we determine the central dark matter density \citep{Read2018}. And~XXV has a low central dark matter density of $\rho_{\rm DM}$(150 pc)= 2.7$^{+1.8}_{-1.6}\times$10$^7$~M$_\odot$ kpc$^{-3}$.}
    \label{fig:mass den prof}
\end{figure}

\subsection{Dark matter density profile of Andromeda XXV} \label{sec:mass profile}
\begin{figure}
	\includegraphics[width=\columnwidth]{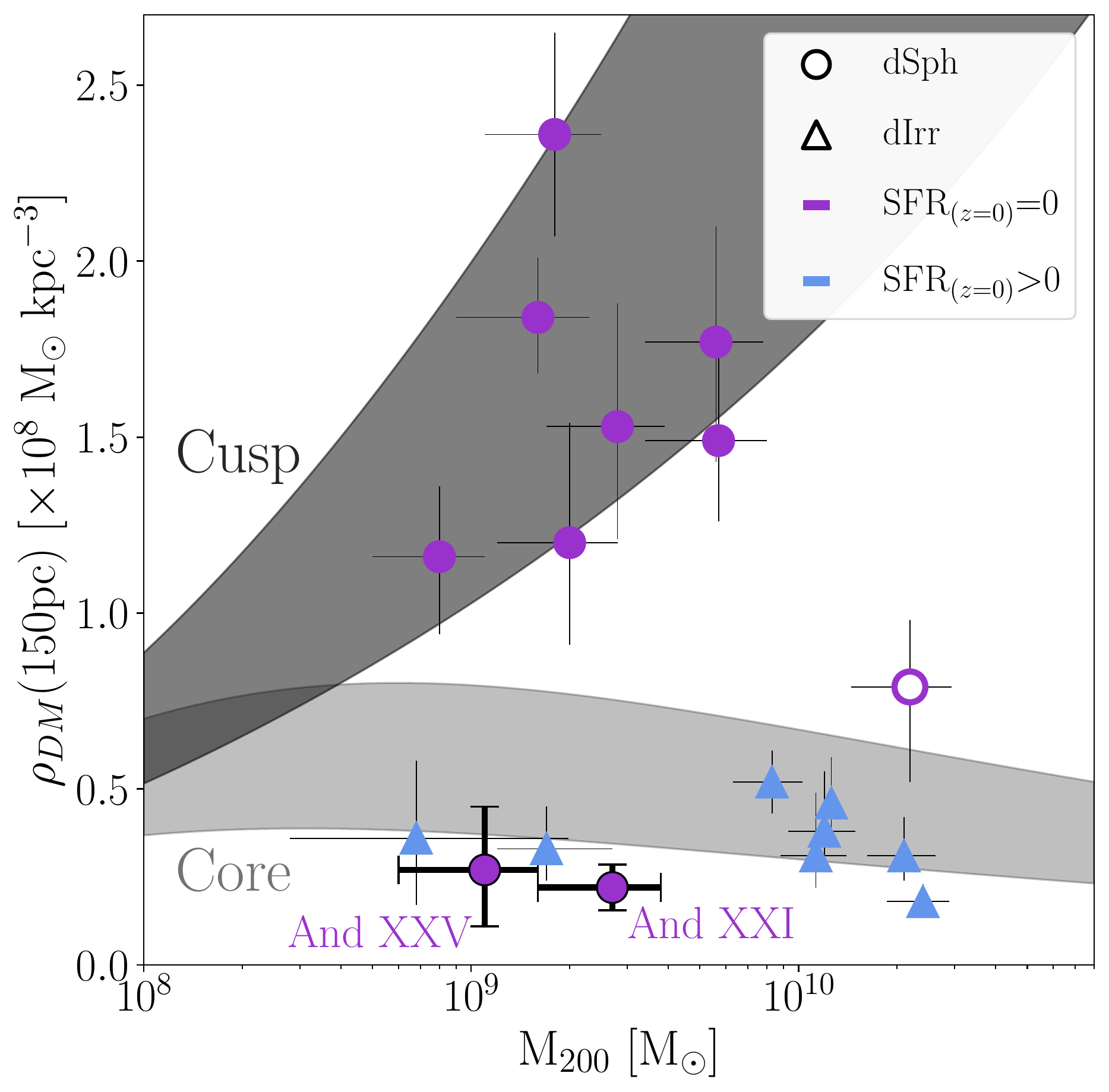}
	\includegraphics[width=\columnwidth]{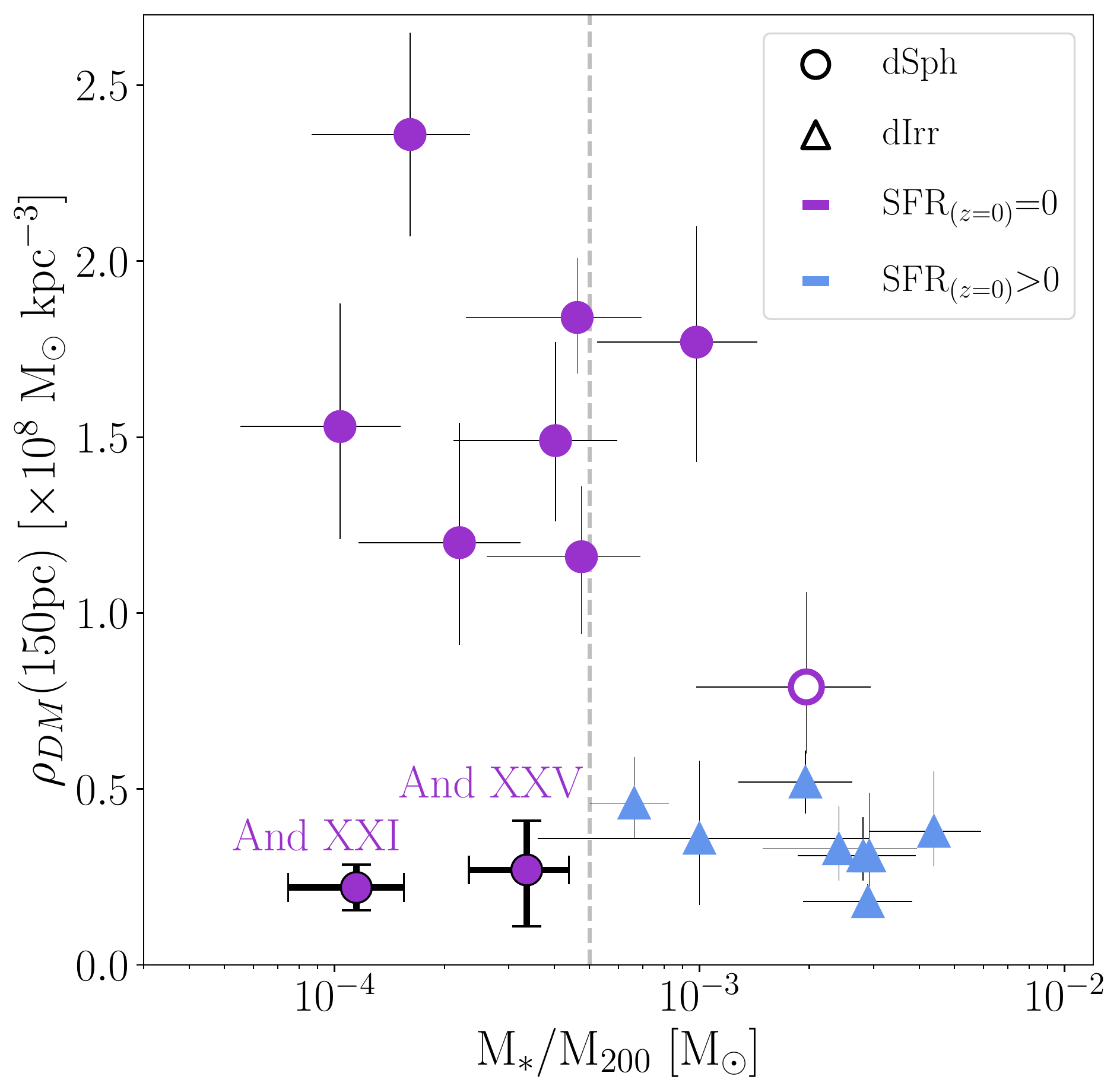}
    \caption{{\bf Top: } The central dark matter density as a function of pre-infall mass, M$_{200}$.
    The dark grey band corresponds to the inner DM densities of completely cuspy profiles, whereas the light grey band corresponds to a fully cored coreNFW profile \citep{Read2016}.
    The width of the bands corresponds to the 1$\sigma$ scatter in DM halo concentrations \citep{Dutton2014}. 
    {\bf Bottom:} Central dark matter density as a function of the stellar mass to halo mass ratio, M$_{\text{*}}$/M$_{200}$. 
    The vertical grey dashed line illustrates the M$_{\text{*}}$/M$_{200}$ ratio above which cusp-core transformations should become efficient \citep{DiCintio2014}. 
    {\bf Both: } Circular data points are dSphs, while triangles are dIrrs. 
    Dwarfs that are currently star forming are light blue whereas dwarfs that are quiescent are purple. 
    Fornax is illustrated by the partially filled purple circular data point due to its unusual star formation history atypical of dSphs.  
    The error bars are the $1\sigma$ uncertainties. 
    And~XXV and And~XXI are labelled and have thicker error bars with caps. 
    Data for And~XXI were obtained from \citet{Collins2021} and from \citet{Read2019} for the other dwarfs.}
    \label{fig:den_m200}
\end{figure}

The resulting dark matter density profile is shown in Fig.~\ref{fig:mass den prof}. The dark grey and light grey shaded regions are the 1 and 2$~\sigma$ uncertainties respectively and the purple dashed line is the half-light radius of And~XXV. Within the uncertainties, it is not possible to distinguish between a cusped or cored profile. Instead, we turn our attention to the central dark matter density, $\rho_{\rm{DM}}(150\rm{pc})$, illustrated by the blue dotted line. This value is used because this is the key region where core formation is expected to reduce dark matter densities, compared to CDM predictions \citep{Read2018, Genina2020b}. For And~XXV we get a central dark matter density of $\rho_{\rm DM}$(150 pc)= 2.7$^{+1.8}_{-1.6}\times$10$^7$~M$_\odot$ kpc$^{-3}$. This is low when compared to other LG dSphs, as shown in Fig.~\ref{fig:den_m200},  with data taken from \cite{Read2019}. Here dSphs are illustrated as circular data points and dIrrs are triangles. Quiescent (non-star forming) dwarfs are purple whereas currently star-forming dwarfs are blue. The sample of dSph galaxies in \citet{Read2019} was chosen to be tidally isolated in order to measure the impact of dark matter heating on $\rho_{\rm DM}$(150 pc), independently of tidal effects. However, while the dSphs in \citet{Read2019} are known to be tidally isolated today (based on their orbits, measured using proper motion data from Gaia+HST \citep{Gaia2016,Gaia2018}, this does not mean that they have been tidally isolated for their whole history. As shown in \citet{Genina2020}, some dwarfs can have past interactions with one another and/or other infalling structure that lowers their inner densities while leaving them on apparently benign orbits today. While it is not possible to control for this for the dSphs in the \cite{Read2019} sample, like Fornax, such effects cannot explain the similarly low density of the isolated dIrrs in the \citet{Read2019} sample. Fornax is illustrated by the partially filled purple circular data point. Fornax is an interesting case that has a density and star formation history similar to dIrrs, but is a dSph. As such, its low density could owe to star formation and/or tides \citep{Genina2022}. A star formation (SF) induced dark matter core provides the best fit to the kinematic data for Fornax, but tides could still have had a major influence even though Fornax’s orbit looks to be tidally benign today.

When comparing And~XXV to the rest of the LG dwarfs we can see just how low And~XXV's central dark matter density is, at approximately an order of magnitude smaller than expected for a quiescent dSph, $\rho_{\rm DM}(150\rm{pc})>10^{8}$~M$_\odot$ kpc$^{-3}$. Instead, And~XXV has a central dark matter density more in line with currently star forming dIrrs, $\rho_{\rm{DM}}(150\rm{pc})<10^{8}$~M$_\odot$ kpc$^{-3}$, which are expected to have lower central dark matter densities due to the dark matter heating effects of their continued star formation. Even when compared to the unusual Fornax, with even lower central density, And~XXV is still a significant outlier. Furthermore, in the top panel in Fig.~\ref{fig:den_m200}, the dark grey band corresponds to a fully cusped profile whereas the light grey band corresponds to a fully cored profile. The width of the bands corresponds to the 1$\sigma$ scatter in DM halo concentrations \citep{Dutton2014}. Here we can see that And~XXV appears to reside in the light grey region which corresponds to a fully cored profile. This behaviour is also reflected in And XXI, another documented M31 outlier for which we have dynamical mass modelling \citep{Collins2021}, as labelled in Fig.~\ref{fig:mass den prof}

\section{Discussion} \label{sec:discussion}
 Dark matter heating is one way in which dwarfs can lower the central dark matter densities. Could the unusually low central density observed in And~XXV be due to dark matter heating caused by extended star formation? From the shallow horizontal branch star formation history, we can see that And~XXV formed 50 per cent of its total stars 8.7 ~Gyrs ago and formed 90 per cent of stars (and thus likely finished star forming) 5.8~Gyrs ago \citep{Weisz2019b}. It is possible that And~XXV had a very short burst of star formation $\sim$1-2 Gyrs ago, although within the uncertainties this is not significant and could be an artefact of the method used as many of the dwarfs in their sample display this short burst. Deeper star formation data is required to confirm the nature of this burst. Dark matter heating is more effective in galaxies with extended star formation \citep{Read2016,Read2019}, hence the relatively short, isolated bursts of star formation history that And~XXV undergoes means it is unlikely that star formation alone could explain the extremely low central density observed in And~XXV. The same conclusion was reached for the other outliers with low central densities \citep{Torrealba2019,Collins2021}. However, in the bottom panel of Fig.~\ref{fig:den_m200} we can see that And~XXV lies below the dashed grey line which indicates the stellar mass-halo mass ratio for which dark matter heating can become viable \citep{DiCintio2014}. However, a recent study by \citet{Orkney2021} showed that gravitational fluctuations due to late minor mergers can also instigate slight core formation, even in dwarfs below this stellar mass-halo mass ratio. Therefore, it is possible that And~XXV may have undergone a slight cusp-core transformation, although none of the simulated, isolated, dwarfs reported in \citet{Orkney2021} reach the low density of And XXV. 

Another process that can lower the central density is tidal interactions. Tidal interactions lower densities at all radii, unlike dark matter heating which only lowers the central density. Tidal interactions can be split into two different processes, tidal stripping and tidal shocking. Tidal stripping occurs when the gravitational force from the host galaxy (in this case M31) exceeds the gravitational force from the dwarf allowing matter (dark matter/stars) to become unbound from the dwarf. Tidal stripping preferentially removes matter from the outer radii working inwards. As such the central dark matter density is only noticeably reduced after significant mass loss. For cusped profiles $\gtrsim 99$~per cent mass of the original mass would need to be lost before the central dark matter density would decrease \citep{Penarrubia2008, Penarrubia2010, Errani2018, Errani2020,Errani2021}. However, cored profiles are less efficient at protecting their central densities and require less extreme mass loss for a noticeable effect \citep{Read2006, Penarrubia2010, Brooks2014}. Tidal shocking is another type of tidal interaction. It occurs for satellites on highly eccentric orbits moving in and out of the gravitational potential of the host. If the resulting gravitational fluctuations occur on shorter time scales than the dynamical time of the interaction, it will dynamically heat the stars and dark matter. The effect of tidal shocking is most pronounced at the pericentre of the orbit, as this is where the gravitational field changes most rapidly. Furthermore, tidal shocking is only effective at lowering central densities for cored systems \citep{Read2006,Errani2017, Errani2018, Errani2020, vandenBosch2018} or for systems that reside in a low concentration dark matter halo \citep{Amorisco2019}. 

At a 3D-project distance of 85.2 kpc \citep{Savino2022}, And~XXV is one of the closest M31 satellites, meaning it may potentially have had a previous encounter with M31 resulting in tidal forces. However, without proper motions, it is difficult to place meaningful constraints on the orbital history to ascertain if And~XXV has had any close approaches with M31. Even armed with hypothetical proper motions, it is important to remember that the current orbital properties are not always a robust indicator of possible past close interactions \citep{Lux2010, Genina2020b}. Without orbital history information, we turn our attention to other potential indicators of tidal interaction. Firstly, we turn to the photometry to see if we can find indications of tidal interactions. No obvious tidal tails or other tidal substructure can be seen in the photometry of And~XXV. The photometry used is shallow, as such tidal features, which tend to have low surface brightnesses, could readily be missed. Without dedicated deep imaging out to the very outskirts of And~XXV, such tidal substructure would be difficult to detect \citep[e.g.][]{Shipp2022}. Recent studies have shown that dSphs can undergo tidal interactions and show no observable tidal substructure \citep{Read2006,Penarrubia2009,Genina2020b}, this is especially true for dwarfs on highly eccentric orbits, since the episodes of shocking are restricted to only the pericentric passages. As such, the dwarfs can relax back to equilibrium on a dynamical timescale, quickly erasing any signs of the interaction from the main body of the dwarf. Moreover, as the stellar component is strongly embedded in the dark matter halo, dSphs can be excessively stripped by tidal forces preferentially removing dark matter and as such, not demonstrate any observable tidal features in the stars until very extreme mass loss \citep[$\gtrsim 90$~per cent][]{Penarrubia2008,Penarrubia2010, Errani2018}. 
Secondly, in the kinematic analysis, we observed a kinematically colder centre in And~XXV with the velocity dispersion increasing with radial distance. This may demonstrate that And~XXV is not in dynamical equilibrium which could be due to previous tidal interactions. Interestingly a similar kinematic profile is observed in Antlia~2, a system for which tides are the suspected culprit \citep{Torrealba2019}. Moreover, the unusually low mass contained within the half-light radius measured for And~XXV could be explained by tidal stripping/shocking, which would reduce the mass of the system over time. This is especially true for cored systems which are able to maintain their original radius while losing mass \citep{Penarrubia2010}, resulting in the extended radial behaviour observed. Finally, dwarfs undergoing strong tidal interactions, which result in stellar mass loss, are expected to be outliers on the luminosity-metallicity relation \citep[e.g. Tucana III, see][]{Simon2017b}. From Fig.~\ref{fig:metal comp} we can see that And~XXV agrees with this relation which suggests it has not lost much stellar mass through tides. However, this does not mean that tides are not important as significant dark matter mass loss can lower the inner density, as can tidal shocking, without any associated stellar mass loss (\citep[e.g.][]{Read2006}).

No one piece of evidence is enough to decisively confirm or omit the possibility of either dark matter heating or tidal interactions as the cause of the low central density. However, the above would imply that we may potentially explain And~XXV's extremely low central density due to a combination of both factors. Meaning And~XXV could have undergone a slight cusp-core transformation from dark matter heating, which made And~XXV more susceptible to tidal stripping and shocking which further reduced the central dark matter density, with tidal interactions being the significant contributing factor. This conclusion is in line with the results from studies investigating the other anomalous dwarfs \citep{Torrealba2019,Collins2021}. Although, detailed analysis in combination with additional data, such as proper motions, is required to consolidate this conclusion. And~XXV joins the small but growing list of unusual LG 'puffy' dwarfs with low central densities. Around the M31 we now see three such systems (And~XXV, And~XIX and And~XXI). A detailed study (Charles et al, in prep) into potential formation and evolution pathways resulting in the low central densities observed will improve our understanding of the nature of dark matter and potentially place constraints on different cosmological models.

\section{Conclusions} \label{sec:conclusions}
We present an updated kinematic analysis for And~XXV. And~XXV was previously identified as a Local Group outlier. Using previous spectroscopic observations in combination with a new data set, providing 53 member stars, more than double that in the previous study, we were sufficiently able to determine the presence of dark matter within And XXV, for the first time. In addition, we dynamically mass model And XXV using \texttt{BINULATOR} + \texttt{GravSphere} to constrain its dark matter density profile. Our key findings are as follows:

\begin{itemize}
  \item We measure the a systemic velocity for And~XXV of $\nu=-107.7\pm{1.0}$~km~s$^{-1}$ and a velocity dispersion of  $\sigma_{\rm\nu}=4.5\pm{1.0}$~km~s$^{-1}$. These values are consistent with the results derived by \citetalias{Collins2013}, which had a much smaller sample size.
  \item We observe that the most central stars have a low velocity dispersion, which increases with increasing radial distance, potentially indicating And~XXV is not in dynamical equilibrium and may have undergone tidal interactions. 
  \item Assuming dynamical equilibrium, we determine the mass contained within the half-light radius to be M(r$<\rm r\textsubscript{h})=6.9^{+3.2}_{-2.8}\times10^6$~M$\textsubscript{\(\odot\)}$, which is lower than we would expect for the size of And~XXV. This mass value corresponds to a mass-to-light ratio of $\text{[M/L]}_{\rm r_{h}}=37^{+17}_{-15}$~M$_\odot$/L$_\odot$,  which, for the first time, indicates And~XXV is a dark matter dominated system. 
  \item For stars with a $S/N>5$ we measure the metallicity of And~XXV to be $\rm[Fe/H]=-1.9\pm{0.1}$~dex from the coadded spectra. We also model the metallicity distribution function from which we are able to resolve a mean metallicity ($\rm[Fe/H]=-2.02\pm0.02$~dex) and metallicity dispersion ($\sigma\textsubscript{[Fe/H]} = 0.08^{+0.02}_{-0.01}$~dex). These results are perfect in agreement with \citetalias{Collins2013} and with the luminosity-metallicity relation for low-mass LG dwarfs \citep{Kirby2013}.
  \item Using the dynamical mass modelling tool \texttt{BINULATOR} + \texttt{GravSphere} we measure a low central dark matter density of $\rho_{\rm DM}$(150 pc)= 2.7$^{+1.8}_{-1.6}\times$10$^7$~M$_\odot$ kpc$^{-3}$. From the dark matter density profile alone we cannot distinguish between a cusped or cored halo. However, when compared to other quiescent star forming dSphs we find the And~XXV has a central dark matter density approximately one order of magnitude smaller than we would expect, more similar to the central dark matter density of isolated star forming dIrrs.
  \item In a companion paper (Charles et al in prep.), we will consider whether dark matter heating, tides, or some combination of these can explain And XXV’s low density in the context of LCDM. We will also explore its implications for alternative dark matter models.
\end{itemize}

\section*{Acknowledgements}
Some of the data presented herein were obtained at the W. M. Keck Observatory, which is operated as a scientific partnership among the California Institute of Technology, the University of California and the National Aeronautics and Space Administration. The Observatory was made possible by the generous financial support of the W. M. Keck Foundation.

The authors wish to recognize and acknowledge the very significant cultural role and reverence that the summit of Mauna Kea has always had within the indigenous Hawaiian community.  We are most fortunate to have the opportunity to conduct observations from this mountain.

The LBT is an international collaboration among institutions in the United States, Italy and Germany. LBT Corporation partners are: The University of Arizona on behalf of the Arizona Board of Regents; Istituto Nazionale di Astrofisica, Italy; LBT Beteiligungsgesellschaft, Germany, representing the Max-Planck Society, The Leibniz Institute for Astrophysics Potsdam, and Heidelberg University; The Ohio State University, representing OSU, University of Notre Dame, University of Minnesota and University of Virginia.

EB acknowledges financial support from a Vici grant from the Netherlands Organisation for Scientific Research (NWO). RI and NM acknowledge funding from the European Research Council (ERC) under the European Unions Horizon 2020 research and innovation programme (grant agreement No. 834148).

\section*{Data Availability}
The raw spectra obtained with DEIMOS are available via the Keck archive. Fully reduced 1D spectra and photometry will be made available upon reasonable request to the lead author. The updated \texttt{GravSphere} code, along with the new BINULATOR binning method is, available to download from \url{https://github.com/justinread/gravsphere}.


\bibliographystyle{mnras}
\bibliography{ref} 



\appendix

\section{Radial Profiles from \texttt{GravSphere}} \label{sec:appendix}
\begin{figure}
	\includegraphics[width=\columnwidth]{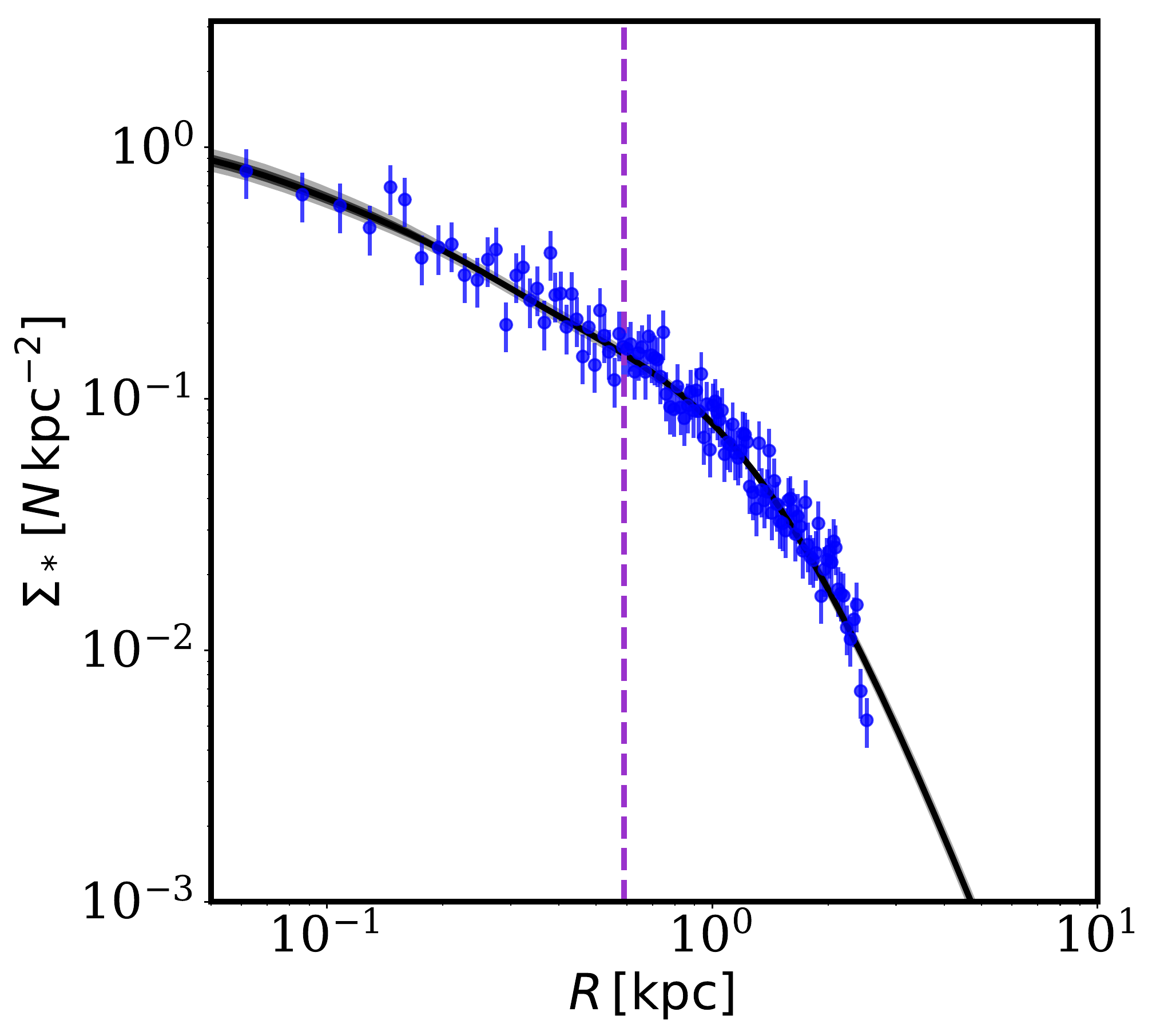}
	\includegraphics[width=\columnwidth]{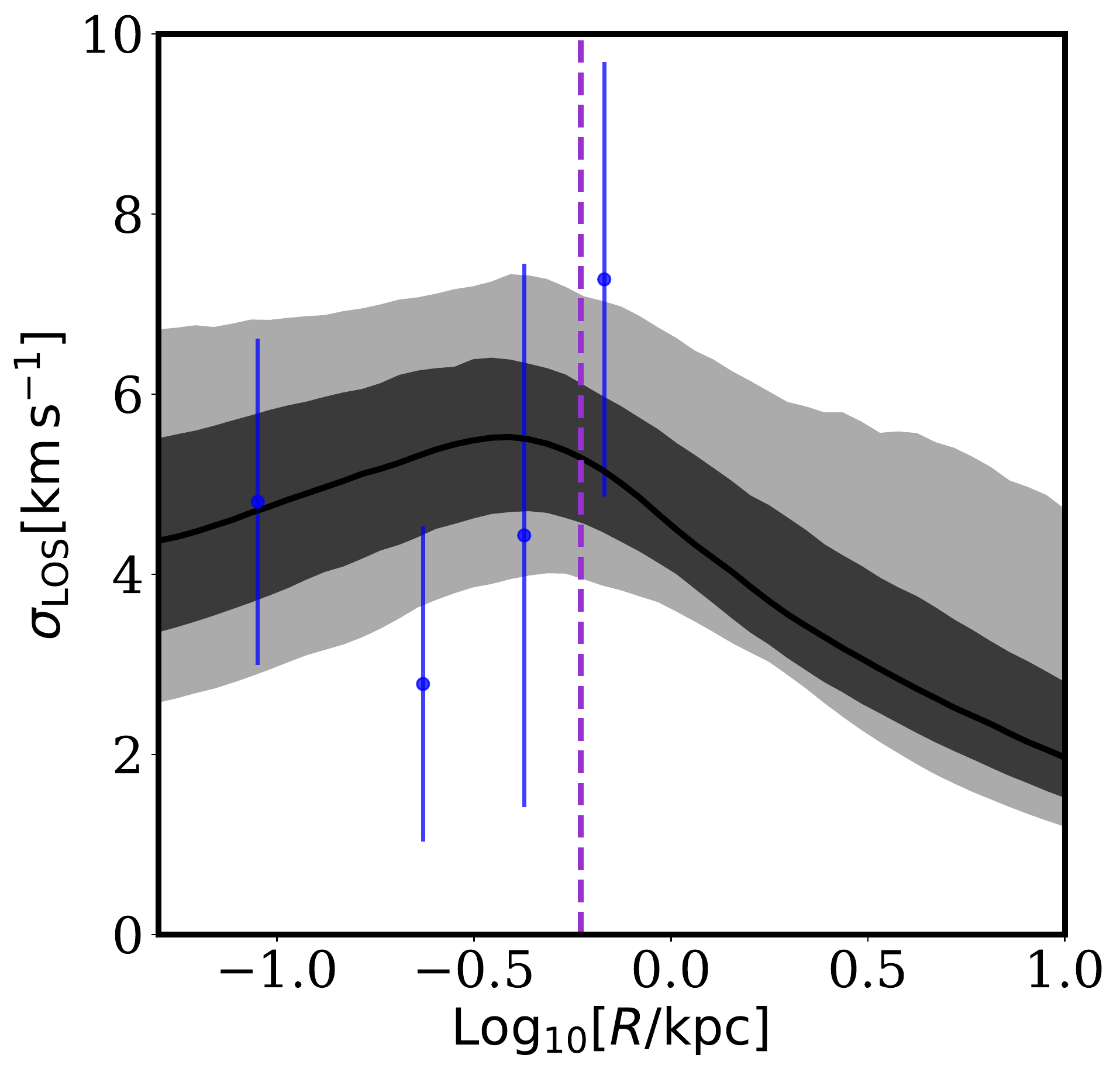}
    \caption{{\bf Top: }The surface brightness profile, $\Sigma_{\text{*}}$. The blue data points are the binned photometry from the LBT imaging data. {\bf Bottom: } The radial velocity dispersion, $\sigma_{\text{LOS}}$. The blue data points are the binned velocity dispersions from our probability weighted member stars. {\bf Both: }The black line shows the fit from \texttt{GravSphere}. The 1 and 2$\sigma$ uncertainty intervals are shown by the dark grey and light grey shaded regions respectively. The vertical purple dashed line is the half-light radius of And~XXV.}
    \label{fig:tracerprof}
\end{figure}

Here we include the radial profile fits from \texttt{GravSphere}, see Fig.~\ref{fig:tracerprof}. First, the top panel shows the well defined surface brightness profile, $\Sigma_{\text{*}}$. The blue data points are the binned photometry from the LBT imaging data out to four effective half light radii. Second, in the bottom panel is the radial velocity dispersion, $\sigma_{\text{LOS}}$. The blue data points are the binned velocity dispersion data taken from our probability weighted member stars. For both panels, the black line shows the fit from \texttt{GravSphere} with the dark and light gray shaded regions showing the 1 and 2$\sigma$ uncertainty intervals respectively. The vertical purple dashed line is the half-light radius of And~XXV. We see that the observed surface brightness profile is well reproduced by \texttt{GravSphere}.  We note that the same velocity dispersion increasing with increasing radius behaviour is observed in the binned kinematic data that is explained in $\S$~\ref{sec:kinematic LG} and $\S$~\ref{sec:discussion}. We reiterate that this behaviour may indicate that And~XXV is not in dynamical equilibrium and could be indicative of And~XXV having undergone tidal interactions. The velocity dispersion profile inferred by GravSphere within the 1$\sigma$ uncertainties for all bins. However, from visual inspection it seems possible that \texttt{GravSphere} has artificially biased the velocity dispersion to higher values at small radii in an attempt to fit this flaring feature. Although, this potential bias would not affect the conclusion that And~XXV has an unusually low central dark matter density. A lower velocity dispersion profile would result in a lower density. As such, the value from this study can be considered an upper bound of the density. 

\bsp	
\label{lastpage}
\end{document}